\documentclass[twocolumn,showpacs,eqsecnum,pra,aps]{revtex4-1}
\usepackage{amsmath}

\usepackage{amssymb}
\usepackage{epsfig}
\usepackage{color}
\begin{document}

\title{Quantum Fisher information on two manifolds of two-mode Gaussian states}
\author{Paulina Marian$^{1,2}$}
\email{paulina.marian@g.unibuc.ro}
\author{ Tudor A. Marian$^{1}$}
\email{tudor.marian@g.unibuc.ro}
\affiliation{ $^1$Centre for Advanced  Quantum Physics,
Department of Physics, University of Bucharest, 
R-077125 Bucharest-M\u{a}gurele, Romania}
\affiliation{$^2$Department of Physical Chemistry, University of Bucharest,\\
Boulevard Regina Elisabeta 4-12, R-030018  Bucharest, Romania}

\date{\today}

\begin{abstract}
We investigate two special classes of two-mode Gaussian states of light that are important
from both the experimental and theoretical points of view: the mode-mixed thermal states 
and the squeezed thermal ones. Aiming to a parallel study, we write the Uhlmann fidelity
between pairs of states belonging to each class in terms of their defining parameters. 
The quantum Fisher information matrices on the corresponding four-dimensional manifolds 
are diagonal and allow insightful parameter estimation. The scalar curvatures 
of the Bures metric on both Riemannian manifolds of special two-mode Gaussian states 
are evaluated and discussed. They are functions of two variables, namely, the mean numbers 
of photons in the incident thermal modes. Our comparative analysis opens the door 
to further investigation of the interplay between geometry and statistics for Gaussian states 
produced in simple optical devices.
\end{abstract}


\maketitle

\section{Introduction}
Answering the question "How close are two states of a quantum system?" is basic 
to the theory of quantum information processing. In principle, the similarity of states 
is decided via quantum-mechanical measurements  whose outcomes are random, 
but have  definite state-depending probability distributions.  There are several 
distance-type measures which are widely used to quantify the probabilistic
distinguishability of two states \cite{NielChua,BZ}. However, in this paper we focus 
on the most convenient of them, namely, the Bures distance that is defined
in connection with quantum fidelity. Besides, the Bures metric for neighbouring 
states is proportional to the quantum Fisher information (QFI) metric 
known to be an essential ingredient in quantum metrology.

On the other hand, our paper deals with two-mode Gaussian states (GSs)
of the quantum radiation field. There are two reasons for this choice.
First, GSs are very useful tools in many quantum optics experiments. 
Some of them are interwoven with quantum information tasks.  As an example, 
we mention that several quantum teleportation experiments were performed 
employing only GSs, starting with the first one carried out 
by Furusawa {\em et al.} \cite{Furusawa}. On the theoretical side, 
the GSs of continuous-variable systems have been intensely investigated 
in the last three decades. Excellent recent reviews cover research on their role
in quantum optics \cite{Olivares} and in quantum information 
as well \cite{WPGCRSL,Adesso}. 

Second, exact explicit formulae have been found for the quantum fidelity 
of arbitrary one-mode \cite{HS1998} and two-mode GSs \cite{PT2012}.
These available compact-form expressions are adequate for various calculations. 
In particular, they allow one a straightforward derivation of the QFI metric 
for the one- and two-mode GSs. In what follows, we address this issue 
by exploiting the fidelity between two-mode GSs. More specifically,
we make a parallel analysis for two classes of such states that are interesting
in both experimental and theoretical research.

Below is presented the outline of the paper in some detail. 
In Sec. II, we start from the analogy between a classical probability distribution 
and that of the outcomes of a quantum-mechanical measurement. 
We recall the definition of the quantum fidelity between two states of a quantum system, 
its connection with their Bures distance, as well as the Uhlmann concise formula.
Emphasis is laid on the proportionality of the infinitesimal Bures distance 
between neighbouring states to their QFI one. In Sec. III,  we  cast the expression 
of the fidelity between two arbitrary  two-mode GSs \cite{PT2012} 
into an alternative form which seems to be more suitable for subsequent calculations.
Section IV examines two important families of two-mode GSs, namely,  
the mode-mixed thermal states (MTSs) and the squeezed thermal states (STSs). 
We briefly review their optical generation, as well as their quantum-mechanical 
description. Despite some formal similarities in their natural parametrizations,
their physical properties are quite different. However, this analysis has lead us to term 
them special two-mode GSs and has suggested that a parallel investigation 
of their closeness features would be appropriate. By applying our alternative 
formula for the fidelity of two-mode GSs, we find in Sec. V the fidelity between MTSs 
and that between STSs, expressed in terms of their specific parameters. 
Section VI is devoted to the derivation of the QFI metric tensors 
on the four-dimensional Riemannian manifolds of the MTSs and STSs. 
Taking advantage that both natural parametrizations are orthogonal ones, 
we evaluate in Sec. VII the scalar curvature of the Bures metric 
on each Riemannian manifold of special two-mode GSs. The scalar 
curvatures of all the MTSs and STSs depend only on their average 
numbers of photons in the incident thermal modes. We explain this specific
property along with an alternative way of deriving both scalar curvatures. 
Their parallel presentation offers the possibility of some interesting comparisons. 
Section VIII summarizes the results and then gives a reliable statistical interpretation
of the Riemannian scalar curvature based on the Bures metric. 
Three appendices detail two basic inequalities involving fidelity in the particular case 
of GSs. Appendix A deals with the inequality between the fidelity and the overlap 
of two $n$-mode GSs. In Appendix B, the property of fidelity of being less than one 
or at most equal to one is checked for thermal states (TSs). Explicit use 
of this conclusion is made in Appendix C in order to verify the same inequality 
for two-mode MTSs and STSs.

\section{Quantum fidelity}

To start on, let $P:=\{ p_b \}, \; (b=1,...,N),$ denote a probability distribution 
assigned to the sample space of an experiment with $N$ outcomes. 
We consider two arbitrary probability distributions,  $P^{\prime}:=\{p_b^{\prime}\}$ 
and $P^{\prime \prime}:=\{p_b^{\prime \prime}\}, $ ascribed to the given sample space. 
Their classical fidelity is the square of the scalar product 
of two unit vectors in $\mathbb{R}^{N}$ with the components
$\{ \sqrt{p_b^{\prime}} \}$ and $\{ \sqrt{p_b^{\prime \prime}} \}$:

\begin{equation}
{\cal F}_{cl}(P^{\prime}, P^{\prime \prime}):
=\left( \sum_{b=1}^N  \sqrt{ p_b^{\prime}\, p_b^{\prime \prime}} \right)^2.
\label{F_c} 
\end{equation}
These vectors define two points on a hyperoctant of the unit sphere $S^{N-1}$ embedded 
in the Euclidean space ${\mathbb R}^{N}.$ 
Their angle at the center of the sphere \cite{Bhatta,Woot}  
is called the Bhattacharyya-Wootters statistical distance \cite{FC}:

\begin{equation}
D_{\rm BW}(P^{\prime}, P^{\prime \prime}):=\arccos{\left( \sqrt{ {\cal F}_{cl} }
(P^{\prime}, P^{\prime \prime} )\right) }.
\label{BW} 
\end{equation}
The square root of the classical fidelity is referred to as affinity of the probability 
distributions  \cite{Luo}. Note that the Hellinger distance between the points 
$ P^{\prime}$ and $ P^{\prime \prime}$  \cite{Luo},
\begin{equation}
D_{\rm H}(P^{\prime}, P^{\prime \prime}):=\left[ \sum_{b=1}^N  \left( \sqrt{ p_b^{\prime}}
-\sqrt{ p_b^{\prime \prime}} \right)^2 \right]^{\frac{1}{2}},
\label{Hell} 
\end{equation}
coincides with their chordal distance and is in turn determined 
by the classical fidelity\ (\ref{F_c}):
\begin{equation}
D_{\rm H}(P^{\prime}, P^{\prime \prime})=\left[ 2-2 \sqrt{ {\cal F}_{cl} }(P^{\prime}, 
P^{\prime \prime})  \right]^{\frac{1}{2}}.
\label{HD} 
\end{equation}

In order to estimate the closeness of two arbitrary states,  
$\hat{\rho}^{\prime}$ and $\hat{\rho}^{\prime \prime}$, 
of a given quantum system, Bures introduced a distance between them  \cite{Bu} 
which is similar to the classical Hellinger distance\ (\ref{HD}):
\begin{equation}
D_{\rm B}(\hat{\rho}^{\prime}, \hat{\rho}^{\prime \prime})
:=\left[ 2-2 \sqrt{ {\cal F}(\hat{\rho}^{\prime}, \hat{\rho}^{\prime \prime})} \right]^{\frac{1}{2}}.
\label{BD} 
\end{equation}
Let  ${\cal H}$  denote the Hilbert space associated to the quantum system. The quantity 
${\cal F}(\hat{\rho}^{\prime}, \hat{\rho}^{\prime \prime})$  occurring in Eq.\ (\ref{BD}) 
is defined as the maximal transition probability between any pair of purifications 
of the states $\hat{\rho}^{\prime}$ and  $\hat{\rho}^{\prime \prime}$. 
However, it is sufficient to restrict our choice of purifications to pairs of state vectors, 
$|{\Psi}^{\prime}\rangle$ and  $|{\Psi}^{\prime \prime}\rangle$, 
both belonging to the tensor product space ${\cal H} \otimes {\cal H}$:

\begin{equation}
{\cal F}(\hat{\rho}^{\prime}, \hat{\rho}^{\prime \prime}):
=\max_{\{|{\Psi}^{\prime}\rangle, |{\Psi}^{\prime \prime}\rangle\}}
|\langle{\Psi}^{\prime}|{\Psi}^{\prime \prime}\rangle|^2.
\label{maxprob} 
\end{equation}
Subsequently \cite{Uhl}, Uhlmann succeeded in deriving the compact expression 
\begin{equation}
{\cal F}(\hat{\rho}^{\prime}, \hat{\rho}^{\prime \prime})=
\left[ {\rm Tr}\left( \sqrt{\sqrt{\hat{\rho}^{\prime \prime}} \hat{\rho}^{\prime}
\sqrt{\hat{\rho}^{\prime \prime}}} \right) \right]^2
\label{F} 
\end{equation}
and interpreted it as a {\em generalization} of the quantum-mechanical 
transition probability between the above states. In a later paper \cite{Jo}, Jozsa coined the name {\em fidelity} 
for the non-negative quantity\ (\ref{F}) and gave elementary proofs 
of its main properties. Note that, when at least one of the quantum states is pure, 
the fidelity\ (\ref{F}) reduces to the Hilbert-Schmidt scalar product of the states:
${\cal F}(\hat{\rho}^{\prime}, \hat{\rho}^{\prime \prime})
={\rm Tr}(\hat{\rho}^{\prime} \hat{\rho}^{\prime \prime})$.
Let us assume, for instance, that $\hat{\rho}^{\prime \prime}$ is a pure state, i. e.,
$\hat{\rho}^{\prime \prime} =|{\psi}^{\prime \prime}\rangle\langle{\psi}^{\prime \prime}|$, 
while the state $\hat{\rho}^{\prime}$ is either pure or mixed. Then their fidelity,
\begin{equation}
{\cal F}(\hat{\rho}^{\prime}, |{\psi}^{\prime \prime}\rangle\langle{\psi}^{\prime \prime}|)
=\langle{\psi}^{\prime \prime}|\hat{\rho}^{\prime}|{\psi}^{\prime \prime}\rangle,
\label{pure} 
\end{equation}
is the probability of transforming the state $\hat{\rho}^{\prime}$ into the state
$\hat{\rho}^{\prime \prime} =|{\psi}^{\prime \prime}\rangle\langle{\psi}^{\prime \prime}|$ 
via a selective measurement. By contrast, the fidelity\ (\ref{F}) has no such operational meaning 
when both states are mixed ones. In this general case, its probabilistic  significance relies 
merely on the Bures-Uhlmann definition\ (\ref{maxprob}).

We mention three properties of the fidelity that are displayed by Eq.\ (\ref{maxprob}).

1) Fidelity is less than one or at most equal to one. This inequality saturates if and only if the states coincide: 
\begin{equation}
{\cal F}({\hat{\rho}}^{\prime}, {\hat{\rho}}^{\prime \prime}) \leqq  1: \quad 
{\cal F}({\hat{\rho}}^{\prime}, {\hat{\rho}}^{\prime \prime}) =1\;    \iff    \; 
{\hat{\rho}}^{\prime \prime}= {\hat{\rho}}^{\prime}.
\label{F<1} 
\end{equation}

2) Symmetry:  
\begin{equation}
{\cal F}({\hat{\rho}}^{\prime \prime}, {\hat{\rho}}^{\prime}) =
{\cal F}({\hat{\rho}}^{\prime}, {\hat{\rho}}^{\prime \prime}).
\label{sym} 
\end{equation}

3) Monotonicity. Fidelity doesn't decrease under any trace-preserving quantum operation 
${\mathcal E}$ performed on both states \cite{NielChua}:
\begin{equation}
{\cal F}({\mathcal E}({\hat{\rho}}^{\prime}), {\mathcal E}({\hat{\rho}}^{\prime \prime})) \geqq
{\cal F}({\hat{\rho}}^{\prime}, {\hat{\rho}}^{\prime \prime}).
\label{mon} 
\end{equation}

Besides, fidelity proved to be an appropriate indicator of the closeness of two quantum states 
via measurements. Recall that for any general measurement there is a Positive Operator-Valued Measure
(POVM)  \cite{NielChua}, i. e., a set of positive operators ${\{\hat E_b\}}$ on the Hilbert space 
${\cal H}$, which provides a resolution of the identity:  ${\sum_{b} \hat E_b}=\hat I$. 
The subscript $b$ indexes the possible outcomes of the measurement whose probabilities 
in the quantum states $\hat{\rho}^{\prime}$ and $\hat{\rho}^{\prime \prime}$ are, respectively, 
$p_{\hat{\rho}^{\prime}}(b)={\rm Tr}(\hat{\rho}^{\prime} \hat E_b)$ 
and $p_{\hat{\rho}^{\prime \prime}}(b)={\rm Tr}(\hat{\rho}^{\prime \prime} \hat E_b).$  
One can extend the formula\ (\ref{F_c}) to define the classical fidelity of these  probability distributions
in a given quantum measurement:
\begin{equation}
{\cal F}_{cl}\left( p_{\hat{\rho}^{\prime}},\, p_{\hat{\rho}^{\prime \prime}};\, \{\hat E_b\}\right):=\left[ \sum_{b}  
\sqrt{p_{\hat{\rho}^{\prime}}(b)\,  p_{\hat{\rho}^{\prime \prime}}(b)} \right]^2.
\label{POVM} 
\end{equation}
In Refs.  \cite{FC,Fuchs,BCFJS} it was  proven an important theorem stating that the quantum 
fidelity is the minimal classical one\ (\ref{POVM}) over the collection of all POVMs:
\begin{equation}
{\cal F}(\hat{\rho}^{\prime}, \hat{\rho}^{\prime \prime})=
\min_{\{\hat E_b\}}\left[ {\cal F}_{cl}\left( p_{\hat{\rho}^{\prime}},\, p_{\hat{\rho}^{\prime \prime}};\,
 \{\hat E_b\}\right) \right]. 
\label{min} 
\end{equation}

When the states $\hat{\rho}^{\prime}$ and $ \hat{\rho}^{\prime \prime}$ commute, then and only then
their fidelity\ (\ref{F}) is equal to the square of their quantum affinity \cite{PT2015}:
\begin{equation}
[\hat{\rho}^{\prime},\,  \hat{\rho}^{\prime \prime}]=\hat 0    \iff   
{\cal F}(\hat{\rho}^{\prime}, \hat{\rho}^{\prime \prime})=
\left[ {\rm Tr}\left( \sqrt{\hat{\rho}^{\prime}}\sqrt{\hat{\rho}^{\prime \prime}} \right) \right]^2.
\label{Fcom} 
\end{equation}
The spectral resolutions of the commuting density operators $\hat{\rho}^{\prime}$ 
and $ \hat{\rho}^{\prime \prime}$ are written in terms of the same complete set 
of orthogonal projections, albeit with specific spectra: 
$\{ {\lambda}_n^{\prime} \}$ and $\{ {\lambda}_n^{\prime \prime} \}$, respectively. 
Therefore, Eq.\ (\ref{Fcom}) takes the form\ (\ref{POVM}): 
\begin{equation}
[\hat{\rho}^{\prime},\,  \hat{\rho}^{\prime \prime}]=\hat 0\;    \iff     \;
{\cal F}(\hat{\rho}^{\prime}, \hat{\rho}^{\prime \prime})=
\left( \sum_{n}  \sqrt{ {\lambda}_n^{\prime} {\lambda}_n^{\prime \prime}}\right)^2.
\label{Fspec} 
\end{equation} 
Accordingly, in the properly termed {\em classical} situation of commuting states, 
the minimum\ (\ref{min}) is reached for a clearly specified projective measurement.

We emphasize that, in the spirit of the correspondence principle, it is possible to {\em guess} the positive operator $\hat{\mathcal B}:
=\sqrt{\hat{\rho}^{\prime \prime}}\hat{\rho}^{\prime}\sqrt{\hat{\rho}^{\prime \prime}}$
\cite{PT2012} occurring in the Uhlmann formula\ (\ref{F}). Indeed, the above structure
of  $\hat{\mathcal B}$ is the simplest one which leads to the classical limit\ (\ref{Fspec}).

Another fidelity-based distance is the Bures angle \cite{NielChua},
\begin{equation}
D_{\rm A}(\hat{\rho}^{\prime}, \hat{\rho}^{\prime \prime}):=\arccos{\left( \sqrt{ {\cal F}(\hat{\rho}^{\prime}, 
\hat{\rho}^{\prime \prime})}\right) }.
\label{D_A} 
\end{equation}
On account of Eqs.\ (\ref{F<1}) and\ (\ref{sym}), both the Bures distance and  the Bures angle 
are genuine metrics since they fulfill in addition the triangle inequality, as shown 
in Refs. \cite{Bu,MNMFL} and, respectively, \cite{NielChua}. Moreover, by virtue 
of the monotonicity property\ (\ref{mon}) of the fidelity, they are contractive distances
(monotone metrics).

Recall that the natural distance between two pure states, 
$|{\psi}^{\prime}\rangle \langle{\psi}^{\prime}|$
and $|{\psi}^{\prime \prime}\rangle \langle{\psi}^{\prime \prime}|$,  
on the manifold of the projective Hilbert space is the Fubini-Study metric \cite{BZ}:
\begin{equation}
D_{\rm FS}{\left( |{\psi}^{\prime}\rangle \langle{\psi}^{\prime}|, |{\psi}^{\prime \prime}\rangle 
\langle{\psi}^{\prime \prime}| \right) }
:=\arccos{\left(|\langle{\psi}^{\prime} |{\psi}^{\prime \prime}\rangle|\right) }.
\label{D_FS} 
\end{equation}
The Bures angle\ (\ref{D_A}) is just the generalization of the Fubini-Study metric 
to the case of mixed states.

We further concentrate on the squared Bures distance between two neighbouring 
quantum states denoted $\hat{\rho}$ and $\hat{\rho}+d\hat{\rho}$:
\begin{equation}
(ds_{\rm B})^2:=[D_{\rm B}(\hat{\rho}, \hat{\rho}+d\hat{\rho})]^2.
\label{ds_B^2} 
\end{equation}
The above squared infinitesimal line element is built with the Bures metric tensor 
on a certain differentiable manifold of quantum states. An important result of Uhlmann's school 
obtained long ago \cite{Hueb} is that the Bures metric\ (\ref{ds_B^2}) is Riemannian.

In a seminal paper  \cite{BC1994}, Braunstein and Caves employ the theory of parameter 
estimation to find the optimal quantum measurement that resolves two neighbouring mixed states. 
Thus they generalize the Wootters statistical distance between pure states \cite{Woot}. 
These authors succeeded in deriving the QFI metric $(ds_{\rm F})^2$ as a reliable 
measure of statistical  distinguishability between neighbouring quantum states. Furthermore, 
a comparison of the QFI metric with H\"ubner's general expression of the infinitesimal 
Bures metric \cite{Hueb} allowed them to establish the basic proportionality formula
\begin{equation}
(ds_{\rm B})^2=\frac{1}{4}(ds_{\rm F})^2.
\label{1:4} 
\end{equation}

A pertinent analysis of the infinitesimal metric on any finite-dimensional state space is carried out 
in Ref. \cite{PeSu} and then carefully reviewed in Ref. \cite{SZ}. It reveals a couple of distinctive 
features of the Bures metric\ (\ref{ds_B^2}). Specifically, this is the minimal one among the monotone,  
Riemannian, and Fisher-adjusted metrics. In addition, it is the only metric from the above class 
whose extension to pure states yields precisely the Fubini-Study metric.

\section{Uhlmann fidelity between two-mode Gaussian states}

In order to tackle the two-mode GSs, we arrange the canonical quadrature  
operators of the modes in a row vector:
\begin{equation}
(\hat u)^T:=(\hat q_1,\; \hat p_1,\; \hat q_2,\; \hat p_2).
\label{u^T}
\end{equation}
Their eigenvalues are the components of an arbitrary dimensionless vector 
$u \in {\mathbb R}^4$: 
\begin{equation}
u^T:=(q_1,\; p_1,\; q_2,\; p_2).
\label{uT}
\end{equation}
Recall that any GS $\hat{\rho}$ is defined by its characteristic function (CF) $\chi (u)$. In turn, 
this is fully determined by the first- and second-order moments of the quadrature operators\ (\ref{u^T})
in the given state $\hat{\rho}$. As a matter of fact, the CF $\chi (u)$ is an exponential 
whose argument is a specific quadratic function of the current vector $u$, Eq.\ (\ref{uT}): 
\begin{equation}
\chi (u)=\exp\left[-\frac{1}{2}\,(J u)^{T}{\mathcal V}\,(J u)+iv^{T}(J u)\right].\label{CF}
\end{equation}
In Eq.\ (\ref{CF}), the components of the vector $v \in \mathbb{R}^4$ are the expectation values 
of the quadrature operators\ (\ref{u^T}) in the chosen GS ${\hat \rho}$:
$v:={\langle{\hat u}\rangle}_{\hat{\rho}}.$
The second-order moments of the deviations from the means 
of the canonical quadrature operators are collected as entries 
of the real and symmetric $4 \times 4$ covariance matrix (CM) 
${\mathcal V}$ of the GS $\hat{\rho}$. In the sequel, we find it often useful
to write the CM partitioned into the following $2\times 2$  submatrices: 
\begin{align}
{\mathcal V}=\left(
\begin{matrix} 
\mathcal V_{1}\;   &   \; {\mathcal C} \\   \\
{\mathcal C}^ T \; &   \;  \mathcal V_{2} 
\end{matrix}
\right).
\label{part}
\end{align} 
The submatrices $\mathcal V_{j}, \;  (j=1,\, 2)$,  are the CMs of the single-mode reduced GSs,
while $ {\mathcal C}$ displays the cross-correlations between the modes.
Further, $J$ denotes the standard  $4\times 4$  matrix of the symplectic form on ${\mathbb R}^4$,
which is block-diagonal and skew-symmetric:
\begin{align}
J:=J_1 \oplus J_2, \quad J_1=J_2:=\left( 
\begin{matrix}
0  & 1\\ -1 & 0
\end{matrix}
\right).
 \label{J}
\end{align}

Any {\em bona fide} CM ${\mathcal V}$  fulfills the concise Robertson-Schr\"odinger 
uncertainty relation:
\begin{equation}
{\zeta}^{\dag}\left({\mathcal V}+\frac{i}{2}J \right){\zeta} \geqq 0, \qquad
\left( \zeta \in \mathbb{C}^4 \right).
\label{R&S}
\end{equation}
Briefly stated, the matrix ${\mathcal V}+\frac{i}{2}J$ has to be positive semidefinite: 
${\mathcal V}+\frac{i}{2}J  \geq 0 $. This requirement is a necessary and sufficient 
condition for the very existence of the Gaussian quantum state 
${\hat \rho}$ \cite{Simon1,HS1989,Simon2}. It implies the inequality 
$\det{\left( {\mathcal V}+\frac{i}{2}J \right) \geqq 0}$
and, in addition, that the CM ${\mathcal V}$ is positive definite: ${\mathcal V}>0$. 
The limit property $\det{\left( {\mathcal V}+\frac{i}{2}J \right) =0}$ is therefore quite special. 
However, it is shared by all the pure GSs,  as well by some interesting mixed ones. 
All these states are said to be at the physicality edge.

In the paper \cite{PT2012} we derived an explicit expression of the fidelity between a pair 
of two-mode GSs, $\hat{\rho}^{\prime}$ and $\hat{\rho}^{\prime \prime}$, with the mean
quadratures $v^{\prime}:={\langle {\hat u} \rangle}_{\hat{\rho}^{\prime}}$ and
$v^{\prime \prime}:={\langle {\hat u} \rangle}_{\hat{\rho}^{\prime \prime}}$, and the CMs
${\mathcal V}^{\prime}$ and ${\mathcal V}^{\prime \prime}$, respectively. 
Let us denote their relative average displacement $\delta v:=v^{\prime}-v^{\prime \prime}$.  
We have found it convenient to employ three determinants satisfying the following inequalities:
\begin{align}
& \Delta:=\det \left({\mathcal V}^{\prime}+{\mathcal V}^{\prime \prime}\right) \geqq 1 \, ;
\label{Delta}  \\
& \Gamma:=2^4{}\det \left[ (J{\mathcal V}^{\prime})
\,(J{\mathcal V}^{\prime \prime})-\frac{1}{4}I\right]  \geqq \Delta \, ; 
\label{Gamma} \\
& \Lambda:=2^4\det \left({\mathcal V}^{\prime }+\frac{i}{2}\,J \right)
\det \left({\mathcal V}^{\prime \prime}+\frac{i}{2}\,J\right) \geqq 0.
\label{Lambda} 
\end{align}
In Eq.\ (\ref{Gamma}),  $I$ denotes the $4 \times 4$ identity matrix. 
The above determinants are manifestly symmetric with respect to the states
$\hat{\rho}^{\prime}$ and $\hat{\rho}^{\prime \prime}$ and so are
the exact expressions of their overlap,
\begin{equation}
{\rm Tr}({\hat{\rho}}^{\prime} {\hat{\rho}}^{\prime \prime} )
=\frac{1}{ \sqrt{\Delta}} 
 \exp{\left[-\frac{1}{2}\left(\delta v \right)^T 
\left({\mathcal V}^{\prime}+{\mathcal V}^{\prime\prime}\right)^{-1} 
\delta v \right]}>0,
\label{overlap}
\end{equation}
and their fidelity \cite{PT2012},
\begin{align}
& {\cal F}({\hat{\rho}}^{\prime}, {\hat{\rho}}^{\prime \prime}) 
= \left[ \left( \sqrt{\Gamma}+\sqrt{\Lambda} \right)
-\sqrt{\left( \sqrt{\Gamma}+\sqrt{\Lambda} \right)^2-\Delta} \right]^{-1} \notag \\
& \times \exp{\left[-\frac{1}{2}\left(\delta v \right)^T 
\left({\mathcal V}^{\prime}+{\mathcal V}^{\prime\prime}\right)^{-1} 
\delta v \right]}.
\label{2F}
\end{align}

It is useful to introduce a pair of non-negative quantities:
\begin{align}
& K_{\pm}:=\sqrt{\Gamma}+\sqrt{\Lambda}\pm \sqrt{\Delta}:  \notag\\
& K_{-} \geqq 0,  \quad  K_{+}-K_{-} \geqq 2.
\label{Kpm}
\end{align}
The proportionality relation\ (\ref{AB}) has the explicit form
\begin{align}
{\cal F}(\hat{\rho}^{\prime}, \hat{\rho}^{\prime \prime})
=\left[ 1+\sqrt{\frac{K_{-}}{\Delta}} \left(  \sqrt{K_{+}}+ \sqrt{K_{-}}\right) \right]
{\rm Tr}({\hat{\rho}}^{\prime} {\hat{\rho}}^{\prime \prime} )>0,
\label{fo2}
\end{align}
which exhibits the general inequality\ (\ref{F>O}). The corresponding
saturation condition,

\begin{align}
& {\cal F}(\hat{\rho}^{\prime}, \hat{\rho}^{\prime \prime})=
{\rm Tr}({\hat{\rho}}^{\prime} {\hat{\rho}}^{\prime \prime} ) \;   \iff   \;  
 K_{-} =0,  \notag\\
 & \text{i. e.,} \;\; \Gamma=\Delta \; \; \text{and} \;\; \Lambda =0,
\label{sat2}
\end{align}
is partly redundant in comparison with the general conclusion\ (\ref{G=D}). Indeed, as shown in Appendix A, 
the latter equality, $\Lambda =0$, is just a consequence of the former, $\Gamma=\Delta$.

We finally write down an alternative form of the fidelity\ (\ref{2F}) that we find appropriate 
to what follows:
\begin{align}
& {\cal F}({\hat{\rho}}^{\prime}, {\hat{\rho}}^{\prime \prime}) 
=2 \left( \sqrt{K_{+}}-\sqrt{K_{-}} \right)^{-2}   \notag\\
& \times \exp{\left[-\frac{1}{2}\left( \delta v \right)^T 
\left( {\mathcal V}^{\prime}+{\mathcal V}^{\prime\prime} \right)^{-1} \delta v \right]}.
\label{2F(K)}
\end{align}

\section{Special two-mode Gaussian states}

We focus on two important families of two-mode GSs that are obtained 
by employing simple optical instruments such as beam splitters 
and non-degenerate parametric down-converters.
Two input light modes interact with the device and their coupling results in two 
output modes \cite{Ulf}. When the incoming beams are chosen to be in TSs,
the outgoing ones are in a two-mode undisplaced GS. In a lossless beam splitter, 
a linear interaction mixes the incident waves to generate a MTS. 
By contrast, in a non-degenerate parametric amplifier, pumping of photons produces 
a non-linear interaction whose outcome is a STS.

Let us summarize the features of preparation and then recall a concise characterization 
of the above output states.

\begin {enumerate}
\item Each incident light wave is in a single-mode TS, so that the global input is their product,
i.e., a two-mode TS on the Hilbert space ${\mathcal H}_1 \otimes {\mathcal H}_2$:
\begin{align}
& \hat{\rho}_{\rm T}(\bar{n}_1, \bar{n}_2):= \left( \hat{\rho}_{\rm T} \right)_1 (\bar{n}_1) 
\otimes \left( \hat{\rho}_{\rm T} \right)_2 (\bar{n}_2):  \notag\\
& \left( \hat{\rho}_{\rm T} \right)_j (\bar{n}_j):=\frac{1}{\bar{n}_j+1}
\exp{\left(  -{\eta}_j {\hat{a}}_j^{\dag} {\hat{a}}_j \right)},  \notag\\ 
& (j=1, 2),
\label{TS} 
\end{align}
with $ {\hat{a}}_j:=\frac{1}{\sqrt{2}}(\hat{q}_j+i\hat{p}_j)$ denoting the photon annihilation 
operator of the mode $j$. In Eq.\ (\ref{TS}), $\bar{n}_j$ is the Bose-Einstein mean photon 
number in the mode $j$,
\begin{equation}
\bar{n}_j=\left[ \exp{({\eta}_j)}-1 \right]^{-1},
\label{BE} 
\end{equation}
and ${\eta}_j$ is the positive dimensionless ratio
\begin{equation}
{\eta}_j:=\frac{\hbar\, {\omega}_j}{k_BT_j}=\ln{\left( \frac{\bar{n}_j+1}{\bar{n}_j} \right) }.
\label{eta} 
\end{equation}

\item The final effect of the mode coupling in both optical devices is modeled by a specific 
unitary operator that induces a linear transformation of the amplitude operators of the modes.

\item It is well known \cite{Duan} that any two-mode GS $\hat{\rho}$ is similar, via a local unitary, 
to another one  whose CM has a unique standard form which consists 
of a partitioning\ (\ref{part}) into special  diagonal $2\times 2$ submatrices:
\begin{align}
& {\mathcal V}_j=b_j{\sigma}_0, \qquad  \left( b_j \geqq \frac{1}{2} \right), \qquad (j=1, 2),  \notag\\
& {\mathcal C}=\left(
\begin{matrix} 
c\;  & 0\; \\ 0\; & d\; 
\end{matrix}
\right),  \qquad  (c \geqq |d| \geqq 0).
 \label{standard}
\end{align}
In Eq.\ (\ref{standard}) and further on,  ${\sigma}_0$ designates the  $2\times 2$ identity matrix. 
The four numbers $b_1, b_2, c,\, \text{and} \; d$ are called the standard-form parameters 
of the given two-mode GS  $\hat{\rho}$. However, there are special classes of two-mode GSs 
with a smaller number of such independent parameters. In particular, for TSs, $c=d=0$, 
while for MTSs, $c=d>0$, and for STSs, $c=-d>0$.
\end {enumerate}

\subsection{Mode-mixed thermal states}

The optical interference of two modes in a reversible, lossless beam splitter is described 
by a mode-mixing operator \cite{Bonny}:
\begin{equation}
\hat M_{12}(\theta,\phi):=\exp{\left[ \frac{\theta}{2}
\left( {\rm e}^{i\phi} \hat a_1  \hat a^{\dag}_2-{\rm e}^{-i\phi} 
\hat a^{\dag}_1  \hat a_2 \right) \right] }.
\label{M_12}
\end{equation} 
Its parameters are the spherical polar angles $\theta$ and 
$\phi: \theta\in [0,\pi),\; \phi\in (-\pi,\pi]$.  The co-latitude $\theta$ determines 
the intensity transmission and reflection coefficients of the device, which are  
$T=\left[ \cos \left( \frac{\theta}{2}\right) \right]^2$
and $R=\left[ \sin \left( \frac{\theta}{2}\right) \right]^2$, respectively.
The longitude $\phi$ accounts for a phase shifting.
As a matter of fact, in view of the Jordan-Schwinger
two-mode bosonic realization of angular momentum \cite{Jordan,Schwinger},
\begin{equation}
\hat{J}_{+}=\hat a^{\dag}_1  \hat a_2, \;\;\;  \hat{J}_{-}= \hat a_1  \hat a^{\dag}_2,  \;\;\,
\hat{J}_3=\frac{1}{2}\left( \hat a^{\dag}_1 \hat a_1-\hat a^{\dag}_2  \hat a_2 \right),
\label{JJJ}
\end{equation} 
the unitary operator\ (\ref{M_12}) is a $SU(2)$ displacement operator \cite{Radcliffe, Arecchi} ,
\begin{equation}
\hat D(\eta):=\exp{\left( \eta\hat{J}_{-}-{\eta}^{\ast} \hat{J}_{+} \right) }, \quad 
\left(  \eta:=\frac{\theta}{2}\, e^{i\phi} \right),
\label{DSU(2)}
\end{equation} 
acting on the two-mode Fock space ${\mathcal H}_1 \otimes {\mathcal H}_2$.
At the same time, we employ the Euler angle parametrization to write it as a  $SU(2)$ 
unitary representation operator whose carrier Hilbert space is  
${\mathcal H}_1 \otimes {\mathcal H}_2$:
\begin{align}
& \hat M_{12}(\theta,\phi)=\hat{\mathcal D} \left[ \, U(\phi, \, \theta, \, -\phi) \, \right]   \notag\\
& =\exp{\left( -i\phi \hat{J}_3 \right) }\, \exp{\left( -i\theta \hat{J}_2 \right) }\, 
\exp{\left( i\phi \hat{J}_3 \right) }.
\label{SU(2)rep}
\end{align} 

When choosing an asymmetrical two-mode TS as input to the beam splitter, 
then we get an emerging MTS as its output:
\begin{equation}
\hat \rho_{\rm MT}=\hat M_{12}(\theta,\phi)\hat \rho_{\rm T}(\bar{n}_1, \bar{n}_2)
\hat M^{\dag}_{12}(\theta,\phi), \;\;  (\bar{n}_1 > \bar{n}_2).
\label{MTS}
\end{equation} 
To the unitary state evolution\ (\ref{MTS}) in the Schr\"{o}dinger picture
is associated the $SU(2)$ matrix $U(\phi, \, \theta, \, -\phi)$ that  transforms 
the annihilation operators in the Heisenberg picture,
\begin{align}
& \left(
\begin{matrix} 
\hat{a}_1^{\prime} \\ \\ \hat{a}_2^{\prime} 
\end{matrix}
\right) 
=U(\phi, \, \theta, \, -\phi)
\left(
\begin{matrix} 
\hat{a}_1 \\ \\ \hat{a}_2
\end{matrix}
\right):    \notag\\   \notag\\
& U(\phi, \, \theta, \, -\phi)=
\left(
\begin{matrix} 
\cos{\left( \frac{\theta}{2} \right) }  &  -\sin{\left( \frac{\theta}{2} \right) }\, e^{-i\phi}  \\
\\ \sin{\left( \frac{\theta}{2} \right) }\, e^{i\phi}  & \cos{\left( \frac{\theta}{2} \right) } 
\end{matrix}
\right).
\label{U}
\end{align}
In turn, the  $SU(2)$ transformation\ (\ref{U}) gives rise to a symplectic orthogonal one of the 
quadratures\ (\ref{u^T}). Its matrix $S(\theta, \phi) \in Sp(4, \mathbb{R}) \cap O(4)$
has the following partition into $2 \times 2$ submatrices:

\begin{align}
S(\theta, \phi)=
\left(
\begin{matrix} 
\cos{\left( \frac{\theta}{2} \right) }\,{\sigma}_0 \ &  -\sin{\left( \frac{\theta}{2} \right) } R(-\phi)  \\
\\ \sin{\left( \frac{\theta}{2} \right) }R(\phi)  & \cos{\left( \frac{\theta}{2} \right) }\, {\sigma}_0 
\end{matrix}
\right).
\label{S(U)}
\end{align}
We have employed the two-dimensional rotation matrix
\begin{align}
& R(\phi):=
\left(
\begin{matrix} 
\cos{\left(\phi  \right)  }\;  &  -\sin{\left( \phi \right) }\,  \\  \\
 \sin{\left( \phi \right) }\;  & \cos{\left( \phi \right) }\,
\end{matrix}
\right),  \quad  (-\pi < \phi \leqq \pi):  \notag\\   \notag\\ 
& R(\phi)=\cos{\left( \phi \right) }\, {\sigma}_0 -i\sin{\left( \phi \right) }\, {\sigma}_2,
\label{R(phi)}
\end{align}
where ${\sigma}_2$ is a Pauli matrix. 
In view of Eq.\ (\ref{VT1}),  the  CM of a two-mode TS\ (\ref{TS}) is diagonal:
\begin{equation}
{\mathcal V}_{\rm T}(\bar{n}_1, \bar{n}_2)= \left( \bar{n}_1+\frac{1}{2} \right) {\sigma}_0 
\oplus \left( \bar{n}_2+\frac{1}{2} \right) {\sigma}_0.
\label{V_in}
\end{equation} 
The unitary similarity\ (\ref{MTS}) of the input and output two-mode GSs is equivalent to 
the symplectic congruence of their CMs:

\begin{align}
& {\mathcal V}_{\rm MT}(\bar{n}_1, \bar{n}_2,  \theta,  \phi)
=S(\theta, \phi)\, {\mathcal V}_{\rm T}(\bar{n}_1, \bar{n}_2)\, S^T(\theta, \phi),    \notag \\  \notag \\
& (\bar{n}_1 > \bar{n}_2).
\label{MTSC}
\end{align}
Accordingly, the CM of the output MTS\ (\ref{MTS}) has the $2 \times 2$ block structure

\begin{align}
{\mathcal V}_{\rm MT}=
\left(
\begin{matrix} 
b_1 {\sigma}_0 \;  &  cR(-\phi) \,  \\  \\
cR(\phi) \;  & b_2 {\sigma}_0 \,
\end{matrix}
\right), 
\label{VMT}
\end{align}
with the standard-form entries:
\begin{align}
& b_{1}= \left( \bar{n}_1+\frac{1}{2} \right) \left[ \cos{ \left( \frac{\theta}{2} \right) } \right]^2
+ \left( \bar{n}_2+\frac{1}{2} \right) \left[ \sin{ \left( \frac{\theta}{2} \right) } \right]^2,  \notag \\
& b_{2}= \left( \bar{n}_1+\frac{1}{2} \right) \left[ \sin{ \left( \frac{\theta}{2} \right) } \right]^2
+ \left( \bar{n}_2+\frac{1}{2} \right) \left[ \cos{ \left( \frac{\theta}{2} \right) } \right]^2,  \notag \\   \notag\\
& c=d=\left( \bar{n}_{1}- \bar{n}_{2}  \right) \cos{\left( \frac{\theta}{2} \right) }
\sin{\left( \frac{\theta}{2} \right) }>0.
\label{sfVMT}
\end{align}
Needless to say,  one gets the standard form\ (\ref{standard}) of the CM  
${\mathcal V}_{\rm MT}(\bar{n}_1, \bar{n}_2, \theta, \phi)$ of a MTS by setting $\phi=0$ 
in Eq.\ (\ref{VMT}). Note also that, in the limit case $ \bar{n}_1= \bar{n}_2=: \bar{n}$,  
a two-mode MTS reduces to the input symmetric two-mode TS with the standard-form 
parameters
\begin{equation}
b_{1}=b_{2}=:b= \bar{n}+\frac{1}{2},     \quad    c=0.
\label{sfTS}
\end{equation}
This happens because the beam splitter has no influence upon two incident light beams
whose one-mode states are identical.

\subsection{Squeezed thermal states}

The coupling of the modes in a non-degenerate parametric amplifier is modelled
by  the action of a two-mode squeeze operator  \cite{SC1985}, 
\begin{align}
& \hat S_{12}(r, \phi):=\exp{\left[ \,r \left( {\rm e}^{i\phi} 
\hat a^{\dag}_1 \hat a^{\dag}_2-{\rm e}^{-i\phi} \hat a_1 
\hat a_2 \right) \right] },   \notag\\
& \left( r>0,\;\;  \phi\in (-\pi,\pi] \right).
\label{S_12}
\end{align} 
The positive dimensionless quantity $r$ is called squeeze parameter  \cite{GK2005}.
Long ago, in a remarkable paper \cite{YCK1986}, Yurke,  McCall, and Klauder introduced
a two-mode bosonic realization of the $su(1,1)$ algebra:
\begin{align}
& \hat{K}_{+}=\hat a^{\dag}_1  \hat a^{\dag}_2,   \quad      \hat{K}_{-}= \hat a_1  \hat a_2,     \notag \\
& \hat{K}_3=\frac{1}{2}\left( \hat a^{\dag}_1 \hat a_1+\hat a_2  \hat a^{\dag}_2 \right) .
\label{KKK}
\end{align}
Starting from these formulae, a $SU(1,1)$ unitary representation on the Hilbert space 
${\mathcal H}_1 \otimes {\mathcal H}_2$ can be decomposed into irreducible unitary
representations of $SU(1,1)$ belonging to the positive discrete series \cite{BL2}.
Note that the Casimir operator
\begin{equation}
\hat{C}:= -\hat{K}_{+}\hat{K}_{-}- \hat{K}_3+ {\hat{K}_3}^2
\label{C}
\end{equation}
and the generator $\hat{K}_3$ are diagonal in the standard Fock basis  
of the Hilbert space ${\mathcal H}_1 \otimes {\mathcal H}_2$. Indeed,
their eigenvalue equations have the solutions:
\begin{align}
& \hat{C}\mid k,m \rangle =k(k-1)\mid k,m \rangle, \quad 
\left( k=\frac{1}{2}, \, 1, \, \frac{3}{2}, \, 2, \, \frac{5}{2}, \dots \right) , \notag\\
& \hat{K}_3\mid k,m \rangle =m\mid k,m \rangle, \quad 
(m=k+l, \;\;  l=0, 1, 2, 3, \dots ): \notag\\
& {\mid k,m \rangle}_{\pm}:=\mid \bar{n}_1, \bar{n}_2 \rangle,  \quad 
k:=\frac{1}{2}\left(  \mid \bar{n}_1- \bar{n}_2 \mid +1 \right), \notag\\
& m:=\frac{1}{2}\left(  \bar{n}_1+ \bar{n}_2 +1 \right), \:
\mid \bar{n}_1- \bar{n}_2 \mid  ={\pm}\left(  \bar{n}_1- \bar{n}_2 \right).
\label{km}
\end{align} 
The two-mode Fock space ${\mathcal H}_1 \otimes {\mathcal H}_2$ is therefore
an orthogonal sum of infinite-dimensional invariant subspaces which are labelled 
with the Bargmann index $k$:

\begin{equation}
{\mathcal H}_1 \otimes {\mathcal H}_2={\mathcal H}^{+\left (\frac{1}{2}\right )} \oplus 
\bigoplus_{k> \frac{1}{2}} \left[ {\mathcal H}_{+}^{+(k)} \oplus {\mathcal H}_{-}^{+(k)} \right]  .
\label{HH}
\end{equation}
This property enables us to write the above-mentioned decomposition of a unitary 
representation of $SU(1,1)$: 

\begin{align}
& \hat{\mathcal D}(V) =\hat {\mathcal D}^{+\left (\frac{1}{2}\right )}(V) \oplus 
\bigoplus_{k> \frac{1}{2}} \left[ \hat{\mathcal D}^{+(k)}(V) \oplus \hat{\mathcal D}^{+(k)}(V)\right] ,  
\notag\\
& \left [ V \in SU(1, 1) \right].
\label{D(V)}
\end{align}

Analogously to  $SU(2)$, the corresponding  $SU(1,1)$ displacement operator acting
on the Hilbert space ${\mathcal H}_1 \otimes {\mathcal H}_2$ \cite{Brif,Novaes},
\begin{align}
& \hat D^{+}(\zeta):=\exp{\left( \zeta \hat{K}_{+}-{\zeta}^{\ast} \hat{K}_{-} \right ) },  \notag\\
& \left(  \zeta:=\frac{\tau}{2}\, e^{i\phi},  \quad  \tau \geqq 0,  \quad -\pi< \phi \leqq \pi \right),
\label{DSU(11)}
\end{align} 
is a  $SU(1,1)$ unitary representation operator as well: 
\begin{align}
& \hat D^{+}(\zeta):= \hat{\mathcal D}{\left[ \, V(\chi, \, \tau, \, -\chi) \, \right]  }  \notag\\
& =\exp{\left( -i\chi \hat{K}_3 \right) }\, \exp{\left( -i\tau \hat{K}_2 \right) }\, 
\exp{\left( i\chi \hat{K}_3 \right) },   \notag\\
& \left( \chi:=-\phi {\pm}\pi: \quad -\pi< \chi \leqq \pi  \right).
\label{DD}
\end{align} 

Owing to the formulae\ (\ref{KKK}), any two-mode squeeze operator\ (\ref{S_12})
is at the same time a  $SU(1,1)$ displacement operator\ (\ref{DSU(11)}) 
with the positive parameter $\tau =2r$:
\begin{equation}
\hat S_{12}(r, \phi)=\hat D^{+}(r\, e^{i\phi}).
\label{S=D}
\end{equation}

When the input to a non-degenerate parametric amplifier is a two-mode thermal
radiation at optical frequencies, then its output is light in a STS:
\begin{equation}
\hat \rho_{\rm ST}=\hat S_{12}(r, \phi)\hat \rho_{\rm T}(\bar{n}_1, \bar{n}_2)
\hat S^{\dag}_{12}(r, \phi).
\label{STS}
\end{equation} 
The unitary transformation\ (\ref{STS}) of the state in the Schr\"{o}dinger picture
determines the $SU(1,1)$ matrix $V(\chi, \, 2r, \, -\chi)$ corresponding 
to a Bogoliubov transformation of the amplitude operators in the Heisenberg picture:
\begin{align}
& \left(
\begin{matrix} 
\hat{a}_1^{\prime} \\  \\ (\hat{a}_2^{\prime})^{\dag} 
\end{matrix}
\right) 
=V(\chi, \, 2r, \, -\chi)
\left(
\begin{matrix} 
\hat{a}_1 \\  \\ {\hat{a}_2}^{\dag}
\end{matrix}
\right):    \notag\\   \notag\\
& V(\chi, \, 2r, \, -\chi)=
\left(
\begin{matrix} 
\cosh(r)  &  \sinh(r) \, e^{i\phi}  \\
\\ \sinh(r) \, e^{-i\phi}  & \cosh(r)
\end{matrix}
\right).
\label{V}
\end{align}
Further, the Bogoliubov transformation\ (\ref{V}) is equivalent to a symplectic one 
of the quadratures\ (\ref{u^T}). Its matrix $S(r,\, \phi) \in Sp(4, \mathbb{R})$
has the following $2 \times 2$ blocks expressed in terms of the identity 
and Pauli matrices:
\begin{align}
& S(r,\, \phi)=    
\left(
\begin{matrix} 
S_a \, &  S_b \,  \\  
S_b \, &  S_a \, 
\end{matrix}
\right):    \quad
S_a:=\cosh(r) {\sigma}_0,       \notag\\  \notag\\ 
& S_b:=  \sinh(r) {\left[ \cos(\phi) {\sigma}_3+ \sin(\phi){\sigma}_1  \right] }.
\label{S(V)}
\end{align}
This symmetric matrix accomplishes a symplectic congruence of the type\ (\ref{MTSC}),
\begin{align}
{\mathcal V}_{\rm ST}(\bar{n}_1, \bar{n}_2, r, \phi)
=S(r, \phi)\: {\mathcal V}_{\rm T}(\bar{n}_1, \bar{n}_2)\: S^T(r, \phi).  
\label{STSC}
\end{align}
We apply Eq.\ (\ref{STSC}) to write the CM ${\mathcal V}_{\rm ST}$ 
of the output STS\ (\ref{STS}). This has the usual partition\ (\ref{part}) 
with the symmetric $2 \times 2$  submatrices:
\begin{align}
& {\mathcal V}_j=b_j{\sigma}_0, \qquad  \left( b_j \geqq \frac{1}{2} \right), \qquad (j=1, 2),  \notag\\  \notag\\
& {\mathcal C}= c\left[ \cos(\phi) {\sigma}_3+ \sin(\phi){\sigma}_1  \right] ,    \qquad (c>0).
\label{VST}
\end{align}
The standard form of the CM ${\mathcal V}_{\rm ST}(\bar{n}_1, \bar{n}_2, r, \phi)$
of a STS is obtained by setting $\phi=0$ in Eq.\ (\ref{VST}) and has the following parameters:
\begin{align}
& b_{1}= \left( \bar{n}_1+\frac{1}{2} \right) \left[ \cosh(r) \right]^2
+ \left( \bar{n}_2+\frac{1}{2} \right) \left[ \sinh(r) \right]^2,   \notag\\
& b_{2}= \left( \bar{n}_1+\frac{1}{2} \right) \left[ \sinh(r) \right]^2
+ \left( \bar{n}_2+\frac{1}{2} \right) \left[ \cosh(r) \right]^2,   \notag\\    \notag\\
& c=-d=\left( \bar{n}_{1}+\bar{n}_{2}+1 \right) \cosh(r) \sinh(r)>0 .
\label{sfVST}
\end{align}

The only pure states belonging to the class of the STSs are the two-mode squeezed 
vacuum states (SVSs). Such a state is the output of a non-degenerate parametric amplifier 
when there is no photon at its input ports, that is, when both incoming field modes 
are in the vacuum state:
\begin{align}
\hat{\rho}_{\rm SV}=|{\Psi}_{\rm SV}\rangle \langle {\Psi}_{\rm SV}| :  \quad
|{\Psi}_{\rm SV}\rangle=\hat S_{12}(r, \phi) | 0,0 \rangle. 
\label{SVS}
\end{align}
Note that the two-mode SVSs make up a two-parameter family of pure symmetric STSs, 
$\left( \bar{n}_{1}=0, \, \bar{n}_{2}=0 \right)$, with the standard-form parameters:
\begin{equation}
b_1=b_2=:b=\frac{1}{2}\cosh(2r), \quad  c=\frac{1}{2}\sinh(2r).
\label{sfVSV}
\end{equation}

We finally mention that a comprehensive study of the transformation of the two-mode GSs 
in a a non-degenerate parametric amplifier, including a detailed analysis 
of the conditions of separability and classicality of the output state, was carried out 
in an earlier paper \cite{PTH2001}. Quite recently, we employed 
the sets of two-mode MTSs and STSs in a comparative investigation 
of the Hellinger distance as a Gaussian measure of all the correlations 
between the modes \cite{PT2015}.

\section{Fidelity between special two-mode Gaussian states}

We consider a pair of special two-mode GSs of the same kind, $\hat{\rho}^{\prime}$ 
and $\hat{\rho}^{\prime \prime}$. Being undisplaced, the states are determined, 
respectively, by their CMs, ${\mathcal V}^{\prime}$ and ${\mathcal V}^{\prime \prime}$, 
whose standard-form parameters are denoted  $\{ b_1^{\prime}, \, b_2^{\prime}, \, c^{\prime}\}$ 
and $\{ b_1^{\prime \prime}, \, b_2^{\prime \prime}, \, c^{\prime \prime}\}$.
At the same time, their fidelity\ (\ref{2F(K)}) has  a simpler form:
\begin{equation}
{\cal F}({\hat{\rho}}^{\prime}, {\hat{\rho}}^{\prime \prime}) 
=2 \left( \sqrt{K_{+}}-\sqrt{K_{-}} \right)^{-2}. 
\label{F2(K)}
\end{equation}

\subsection{Mode-mixed thermal states}

Let $\{ \bar{n}_1^{\prime}, \, \bar{n}_2^{\prime}, \, {\theta}^{\prime}, \, {\phi}^{\prime}\} \,$
and $\{ \bar{n}_1^{\prime \prime}, \, \bar{n}_2^{\prime \prime}, \, {\theta}^{\prime \prime}, \, 
{\phi}^{\prime \prime}\} \,$ stand for the parameters of the MTSs  $\hat{\rho}^{\prime}$ 
and $\hat{\rho}^{\prime \prime}$, respectively. Making use of the CM\ (\ref{VMT}), 
we have evaluated the determinants $\Delta$, Eq.\ (\ref{Delta}) 
and $\Gamma$, Eq.\ (\ref{Gamma}), via the partitions of the corresponding 
$4\times 4$ matrices into $2\times 2$ blocks. We have applied the Schur determinant 
factorization (as the product of the determinant of a $2\times 2$ principal submatrix 
by that of its Schur complement) to obtain the formulae:

\begin{align} 
& \Delta=\left\{ \left( b^{\prime}_{1}+b^{\prime \prime}_{1} \right)
\left( b^{\prime}_{2}+b^{\prime \prime}_{2} \right) \right.   \notag \\
& \left. -\left[ \left( c^{\prime} \right)^2+\left( c^{\prime \prime} \right)^2
+2c^{\prime}c^{\prime \prime}\cos{\left( {\phi}^{\prime}-{\phi}^{\prime \prime} \right) }\right] 
 \right\}^2 ;         
\label{DeltaM}
\end{align}
\begin{align} 
& \Gamma =16\left\{ \left[ b^{\prime}_{1}b^{\prime}_{2}-\left( c^{\prime} \right)^2 \right] 
\left[ b^{\prime \prime}_{1}b^{\prime \prime}_{2}-\left( c^{\prime \prime} \right)^2 \right] \right.  \notag \\
& \left. +\frac{1}{4}\left[ b^{\prime}_{1}b^{\prime \prime}_{1}
+b^{\prime}_{2}b^{\prime \prime}_{2}+2c^{\prime}c^{\prime \prime} 
\cos{\left( {\phi}^{\prime}-{\phi}^{\prime \prime} \right) }\right] +\frac{1}{16} \right\}^2.
\label{GammaM}
\end{align}
The  determinant $\Lambda$ is the product\ (\ref{Lambda}) of two similar symplectic invariants:
\begin{align} 
& \Lambda=16\left\{  \left[ b^{\prime}_{1}b^{\prime}_{2}-\left( c^{\prime}\right)^2 \right]^2  \right.  \notag \\
&  \left. -\frac{1}{4}\left[ \left( b_{1}^{\prime}\right)^2+\left( b_{2}^{\prime}\right)^2
+2\left( c^{\prime}\right)^2 \right]+\frac{1}{16} \right\}  
\left\{ \left[ b^{\prime \prime}_{1}b^{\prime \prime}_{2}-\left( c^{\prime \prime}\right)^2 \right]^2  \right.  \notag \\
&  \left. -\frac{1}{4}\left[ \left( b_{1}^{\prime \prime}\right)^2+\left( b_{2}^{\prime \prime}\right)^2
+2\left( c^{\prime \prime}\right)^2 \right]+\frac{1}{16} \right\}.
\label{LambdaM}                                   
\end{align}
In the resulting functions $K_{\pm}$, Eq.\ (\ref{Kpm}), we substitute the specific 
expressions\ (\ref{sfVMT}) of the standard-form parameters and get the following couple of formulae:

\begin{align} 
& K_{+}=2\left\{ \left(  \bar{n}_1^{\prime} \, \bar{n}_2^{\prime} \, \bar{n}_1^{\prime \prime} \, 
\bar{n}_2^{\prime \prime} \, \right)^{\frac{1}{2}} \right.  \notag \\
& \left. +\left[ \left(  \bar{n}_1^{\prime}+1 \right) \left( \bar{n}_2^{\prime}+1 \right)
\left( \bar{n}_1^{\prime \prime}+1 \right) \left( \bar{n}_2^{\prime \prime}+1 \right) \right]^{\frac{1}{2}} 
\right\}^2.   
\label{K(+)M}
\end{align}
\begin{align} 
& K_{-}=2\left\{ \left[  \bar{n}_1^{\prime} \left( \bar{n}_2^{\prime}+1 \right)  \bar{n}_1^{\prime \prime}  
\left( \bar{n}_2^{\prime \prime}+1 \right) \right]^{\frac{1}{2}} \right.  \notag \\
& \left. +\left[ \left(  \bar{n}_1^{\prime}+1 \right)  \bar{n}_2^{\prime}
\left( \bar{n}_1^{\prime \prime}+1 \right) \bar{n}_2^{\prime \prime} \right]^{\frac{1}{2}} \right\}^2   \notag \\ 
& -\left( \bar{n}_1^{\prime}-\bar{n}_2^{\prime} \right)\, \left( \bar{n}_1^{\prime \prime}
-\bar{n}_2^{\prime \prime} \right) \left\{ 1-\cos{\left( {\theta}^{\prime}-{\theta}^{\prime \prime} \right) }\right.  \notag \\ 
& \left. +\sin{\left( {\theta}^{\prime}  \right) } \sin{\left( {\theta}^{\prime \prime} \right) }
\left[ 1-\cos{\left( {\phi}^{\prime}-{\phi}^{\prime \prime} \right) }\right]  \right\} . 
\label{K(-)M}
\end{align}
Insertion of Eqs.\ (\ref{K(+)M}) and\ (\ref{K(-)M}) into Eq.\ (\ref{F2(K)}) gives the fidelity of two MTSs.
When all the other parameters of both states are kept fixed, this fidelity  is an even function
of the phase difference ${\phi}^{\prime}-{\phi}^{\prime \prime}$, which is strictly decreasing 
in the interval $[0, \pi]$.

\subsection{Squeezed thermal states}

We focus on a pair of  STSs, $\hat{\rho}^{\prime}$ and $\hat{\rho}^{\prime \prime}$,  
and designate their sets of parameters as $\{ \bar{n}_1^{\prime}, \, \bar{n}_2^{\prime}, \, r^{\prime}, 
\, {\phi}^{\prime} \} \,$ and, respectively, $\{ \bar{n}_1^{\prime \prime}, \, \bar{n}_2^{\prime \prime}, \, 
r^{\prime \prime}, \, {\phi}^{\prime \prime} \} \,$. Starting from the CM 
${\mathcal V}_{\rm ST}(\bar{n}_1, \bar{n}_2, r, \phi)$, specified by Eqs.\ (\ref{part}) and\ (\ref{VST}),
and employing the same technique as for MTSs, we have evaluated 
the determinants\ (\ref{Delta})-\ (\ref{Lambda}):

\begin{align} 
& \Delta=\left\{ \left( b^{\prime}_{1}+b^{\prime \prime}_{1} \right)
\left( b^{\prime}_{2}+b^{\prime \prime}_{2} \right) \right.   \notag \\
& \left. -\left[ \left( c^{\prime} \right)^2+\left( c^{\prime \prime} \right)^2
+2c^{\prime}c^{\prime \prime}\cos{\left( {\phi}^{\prime}-{\phi}^{\prime \prime} \right) }\right] 
 \right\}^2 ;         
\label{DeltaS}
\end{align}
\begin{align} 
& \Gamma =16\left\{ \left[ b^{\prime}_{1}b^{\prime}_{2}-\left( c^{\prime} \right)^2 \right] 
\left[ b^{\prime \prime}_{1}b^{\prime \prime}_{2}-\left( c^{\prime \prime} \right)^2 \right] \right.  \notag \\
& \left. +\frac{1}{4}\left[ b^{\prime}_{1}b^{\prime \prime}_{1}
+b^{\prime}_{2}b^{\prime \prime}_{2}-2c^{\prime}c^{\prime \prime} 
\cos{\left( {\phi}^{\prime}-{\phi}^{\prime \prime} \right) }\right] +\frac{1}{16} \right\}^2;
\label{GammaS}
\end{align}
\begin{align} 
& \Lambda=16\left\{  \left[ b^{\prime}_{1}b^{\prime}_{2}-\left( c^{\prime}\right)^2 \right]^2
-\frac{1}{4}\left[ \left( b_{1}^{\prime}\right)^2+\left( b_{2}^{\prime}\right)^2
-2\left( c^{\prime}\right)^2 \right]+\frac{1}{16} \right\}  \notag \\
& \times \left\{ \left[ b^{\prime \prime}_{1}b^{\prime \prime}_{2}-\left( c^{\prime \prime}\right)^2 \right]^2 
-\frac{1}{4}\left[ \left( b_{1}^{\prime \prime}\right)^2+\left( b_{2}^{\prime \prime}\right)^2
-2\left( c^{\prime \prime}\right)^2 \right]+\frac{1}{16} \right\}.
\label{LambdaS}                                   
\end{align}
Substitution of the above formulae into Eq.\ (\ref{Kpm}) and subsequent insertion of the specific 
standard-form parameters\ (\ref{sfVST}) yield the following functions:

\begin{align}
& K_{+}=2\left\{ \left[ \bar{n}_1^{\prime} \, \bar{n}_2^{\prime}\left(  \bar{n}_1^{\prime \prime} +1 \right)  
\left( \bar{n}_2^{\prime \prime}+1 \right) \right]^{\frac{1}{2}} \right.  \notag \\
& \left. +\left[ \left(  \bar{n}_1^{\prime}+1 \right) \left( \bar{n}_2^{\prime}+1 \right)
\bar{n}_1^{\prime \prime} \, \bar{n}_2^{\prime \prime} \right]^{\frac{1}{2}} \right\}^2   \notag \\ 
& +\left( \bar{n}_1^{\prime}+\bar{n}_2^{\prime}+1 \right) \left( \bar{n}_1^{\prime \prime}
+\bar{n}_2^{\prime \prime}+1 \right) \left\{ 1+\cosh{\left[ 2\left( r^{\prime}-r^{\prime \prime} \right) \right] }\right.  
\notag \\ 
& \left. +\sinh{\left( 2r^{\prime}  \right) } \sinh{\left( 2r^{\prime \prime} \right) }
\left[ 1-\cos{\left( {\phi}^{\prime}-{\phi}^{\prime \prime} \right) }\right]  \right\} . 
\label{K(+)S}
\end{align}

\begin{align} 
& K_{-}=2\left\{ \left[  \bar{n}_1^{\prime} \left( \bar{n}_2^{\prime} +1 \right) \bar{n}_1^{\prime \prime} 
\left( \bar{n}_2^{\prime \prime} +1 \right) \right]^{\frac{1}{2}} \right.  \notag \\
& \left. +\left[ \left(  \bar{n}_1^{\prime}+1 \right) \bar{n}_2^{\prime}
\left( \bar{n}_1^{\prime \prime}+1 \right)  \bar{n}_2^{\prime \prime} \right]^{\frac{1}{2}} 
\right\}^2.   
\label{K(-)S}
\end{align}
By introducing the functions\ (\ref{K(+)S}) and\ (\ref{K(-)S}) into Eq.\ (\ref{F2(K)}),  we recover
the expression of the fidelity between two STSs that has previously been used to quantify 
the Gaussian entanglement of such a two-mode state \cite{PTH2003}. Concerning its dependence 
on the phase difference ${\phi}^{\prime}-{\phi}^{\prime \prime}$, the fidelity between two STSs 
is an even function of this variable and strictly decreases with  it in the interval $[0, \pi]$.

\section{Quantum Fisher information tensor on the manifolds of special two-mode Gaussian states}

Let us look at a quantum system which has a manifold of states that are characterized 
by a finite set of continuous real variables $\{ \xi \}$. We concentrate on a pair of neighbouring
states,  $\hat{\rho}(\xi)$ and $\hat{\rho}(\xi+t d\xi)$, where $t$ is a real non-negative variable.
We apply Eq.\ (\ref{BD}):          
\begin{align}
& \frac{1}{2}\left\{ D_{\rm B}{ \left[ \hat{\rho}(\xi), \,\hat{\rho}(\xi+t d\xi) \right] } \right\}^2
=1-\sqrt{ {\cal F}(t) }:    \notag \\  
& {\cal F}(t):={\cal F}{  \left[ \hat{\rho}(\xi), \, \hat{\rho}(\xi+t d\xi) \right]  },   \quad
\left( t \geqq 0 \right),
\label{BDn}
\end{align}
Note the general properties:
\begin{equation}
{\cal F}(0)=1,  \qquad   -\left[  \frac{d}{dt}\sqrt{ {\cal F}(t) } \right]_{t=0}=0. 
\label{F(0)}
\end{equation}
The former identity in Eq.\ (\ref{F(0)}) represents the sufficiency part of the saturation case
in Eq.\ (\ref{F<1}),  while the latter was proven by H\"{u}bner in Ref. \cite{Hueb}. 
Therefore, the first non-vanishing term in the Maclaurin series 
of the squared Bures distance\ (\ref{BDn}) is the $t^2$ term:
\begin{equation}
\left\{ D_{\rm B}{ \left[ \hat{\rho}(\xi), \,\hat{\rho}(\xi+t d\xi) \right] } \right\}^2
=t^2 \left( ds_{\rm B} \right)^2+{\rm O}(t^3).
\label{t^2}
\end{equation}
Its coefficient is the squared infinitesimal Bures line element on the above-specified manifold,
\begin{equation}
\left( ds_{\rm B} \right)^2= \sum_{\alpha} \sum_{\beta} \,  g_{\alpha \beta}(\xi) 
d{\xi}_{\alpha}  d{\xi}_{\beta},
\label{u^2}
\end{equation}
where $g_{\alpha \beta}(\xi)$ are the components of the affiliated Riemannian metric tensor.
We will evaluate it for two-mode MTSs and STSs as the second-order derivative
\begin{equation}
\left( ds_{\rm B} \right)^2=-\left[  \frac{d^2}{dt^2}\sqrt{ {\cal F}(t) } \right]_{t=0} \geqq 0. 
\label{dsB}
\end{equation}
In the realm of the GSs, this method was first applied by Twamley to evaluate the Bures 
geodesic metric for one-mode STSs \cite{Twamley}. Then it was used to evaluating 
and studying the QFI metric for single-mode displaced TSs \cite{PS1998} and quite recently
for arbitrary one-mode GSs \cite{Pinel}. Here we take advantage of the natural parametrizations 
for both families of two-mode states, MTSs and STSs, as well as of the convenient formulae 
for their fidelity established in Sec. V.
 
\subsection{Mode-mixed thermal states}

The fidelity\ (\ref{BDn}) between neighbouring MTSs, 
\begin{align}
& {\cal F}(t):={\cal F} \left[ \hat{\rho} \left(  \bar{n}_1, \, \bar{n}_2, \, \theta, \, \phi \right), \right.  \notag \\
& \left. \hat{\rho} \left(  \bar{n}_1+t d \bar{n}_1, \, \bar{n}_2+t d \bar{n}_2, \,    
\theta+t d\theta, \, \phi+t d\phi \right)  \right],     \notag \\
& \left( t \geqq 0 \right),
\label{FnMTS}
\end{align}
has the expression\ (\ref{F2(K)}),
\begin{equation}
{\cal F}(t)=2 \left[ \sqrt{K_{+}(t)}-\sqrt{K_{-}(t)} \right]^{-2}, 
\label{F2(Kt)}
\end{equation}
with the functions $K_{\pm}(t)$ obviously introduced in the manner of Eq.\ (\ref{FnMTS}).
Hence the Bures metric\ (\ref{dsB})  for MTSs reads:

\begin{equation}
\left( ds_{\rm B} \right)^2=-\sqrt{2}\left\{ \frac{d^2}{dt^2}\left[ \sqrt{K_{+}(t)}
-\sqrt{K_{-}(t)} \right]^{-1} \right\}_{t=0}. 
\label{dsSG}
\end{equation}
One can readily check the identities\ (\ref{F(0)}), the first one being precisely the sufficient condition
in Eq.\ (\ref{FMT=1}). A straightforward  calculation then yields the formula:
\begin{align}
& \left[ \left( ds_{\rm B} \right)_{\rm MT} \right]^2=\frac{1}{4}\left( \frac{1}{\bar{n}_1(\bar{n}_1+1)}
\left( d\bar{n}_1 \right)^2
+\frac{1}{\bar{n}_2(\bar{n}_2+1)}\left( d\bar{n}_2 \right)^2   \right.  \notag \\     
& \left. +\frac{\left( \bar{n}_1-\bar{n}_2 \right)^2 }{2\bar{n}_1 \bar{n}_2+ \bar{n}_1+\bar{n}_2 }   
 \left\{ (d\theta)^2 +\left[ \sin(\theta) \right]^2  (d\phi)^2 \right\} \right).
\label{dsMT}
\end{align}
We have thus checked that the differentiable manifold 
${\mathcal M} {\left(  \bar{n}_1, \, \bar{n}_2, \, \theta, \, \phi \right)}$ 
of the two-mode MTSs equipped with the Bures metric\ (\ref{dsMT}) is a Riemannian one. 
Besides, its metric tensor has a diagonal matrix 
$g_{\rm MT}{\left(  \bar{n}_1, \, \bar{n}_2, \, \theta, \, \phi \right)}$. 
Because the first two terms in Eq.\ (\ref{dsMT}) are not influenced by the action 
of the beam splitter, they depend only on the input two-mode TS
$\hat \rho_{\rm T}(\bar{n}_1, \bar{n}_2)$. Their sum defines therefore the Bures metric 
on the two-dimensional manifold  ${\mathcal M} {\left(  \bar{n}_1, \, \bar{n}_2 \right) }$
of the two-mode TSs\ (\ref{TS}): 
\begin{align}
& \left[ \left( ds_{\rm B} \right)_{\rm T} \right ]^2=\frac{1}{4}\left[ \frac{1}{\bar{n}_1(\bar{n}_1+1)}
\left( d\bar{n}_1 \right)^2  \right.  \notag \\   
& \left. +\frac{1}{\bar{n}_2(\bar{n}_2+1)}\left( d\bar{n}_2 \right)^2  \right], 
\label{dsT}
\end{align}
With the reparametrization $\sqrt{\bar{n}_j}:=\sinh(x_j), \, (j=1,2), $
the metric\ (\ref{dsT}) becomes an Euclidean flat one:
\begin{equation} 
\left[ \left( ds_{\rm B} \right)_{\rm T} \right ]^2=(dx_1)^2+(dx_2)^2. 
\label{dsTR^2}
\end{equation} 
This means that the Riemannian manifold 
${\mathcal M} {\left(  \bar{n}_1, \, \bar{n}_2 \right) }$ is locally isometric with the first quadrant 
$\mathbb{R}_{+}^2$ of the Euclidean plane. 
The last two terms in Eq.\ (\ref{dsMT}) originate in the interaction of the incoming thermal modes 
with the beam splitter resulting in the $SU(2)$ unitary state evolution\ (\ref{MTS}). 
Their sum is proportional to the Euclidean round metric 
on the two-dimensional unit sphere $S^2$:
\begin{equation}
\left( ds_{\theta,  \phi} \right)^2= (d\theta)^2 \, +\left[ \sin(\theta) \right]^2  (d\phi)^2.
\label{S2}
\end{equation}
In this line, $S^2$ can be viewed as a compact homogeneous space: $S^2=SU(2)/U(1)$.

Owing to the general relation\ (\ref{1:4}),  Eq.\ (\ref{dsMT}) provides additionally the statistical
distance
\begin{align}
& \left[ \left( ds_{\rm F} \right)_{\rm MT} \right]^2=H_{\bar{n}_1} \left( d\bar{n}_1 \right)^2
+H_{\bar{n}_2} \left( d\bar{n}_2 \right)^2    \notag \\     
& +H_{\theta} (d\theta)^2 +H_{\phi} (d\phi)^2.
\label{dsFMT}
\end{align}
The components of  the diagonal QFI metric tensor are independent of the phase $\phi$: 
\begin{align}
& H_{\bar{n}_1}=\frac{1}{\bar{n}_1(\bar{n}_1+1)},     \qquad
H_{\bar{n}_2}=\frac{1}{\bar{n}_2 (\bar{n}_2+1)},         \notag \\     
& H_{\theta}=\frac{\left( \bar{n}_1-\bar{n}_2 \right)^2 }
{2\bar{n}_1 \bar{n}_2+ \bar{n}_1+\bar{n}_2 },  \notag \\ 
& H_{\phi}=\frac{\left( \bar{n}_1-\bar{n}_2 \right)^2 \, \left[ \sin(\theta) \right]^2 }
{2\bar{n}_1 \bar{n}_2+ \bar{n}_1+\bar{n}_2 }.
\label{HMT}
\end{align}
Since the above QFI matrix is diagonal,  the natural parameters 
$\left\{  \bar{n}_1, \, \bar{n}_2, \, \theta, \, \phi \right \} $ of the MTSs are said to be
orthogonal.  According to Eqs.\ (\ref{dsFMT}) and\ (\ref{HMT}), 
the quantum Cram\'{e}r-Rao lower bound for the variance of such a state estimator 
${\xi}_{\alpha}$  reads \cite{Paris}:
\begin{equation}
\left( \Delta {\xi}_{\alpha} \right)^2 \geqq \frac{1}{{\cal N} H_{{\xi}_{\alpha}}},   \qquad
({\xi}_{\alpha} =\bar{n}_1, \, \bar{n}_2, \, \theta, \, \phi),  
\label{CRBM}
\end{equation}
where  ${\cal N}$ is the number of measurements.

To sum up, the Bures metric\ (\ref{dsMT}) has the structure:
\begin{align} 
& \left[ \left( ds_{\rm B} \right)_{\rm MT} \right]^2=\left[ \left( ds_{\rm B} \right)_{\rm T} \right ]^2
+\left[ f_{\rm MT}{\left(  \bar{n}_1, \bar{n}_2 \right) } \right]^2 \left( ds_{\theta,  \phi} \right)^2:   \notag \\
& f_{\rm MT}{\left(  \bar{n}_1, \bar{n}_2 \right) }:=\frac{1}{2}\sqrt{H_{\theta}}.
\label{dsMT+}
\end{align}
In addition, let us write down the Bures-metric volume element 
on the  Riemannian manifold of the two-mode MTSs:
\begin{equation}
d{\mathcal V}_{\rm B}:=\sqrt{\det{\left[ g_{\rm MT}
{\left(  \bar{n}_1, \, \bar{n}_2, \, \theta, \, \phi \right)} \right] }} \,
d\bar{n}_1 \, d\bar{n}_2 \, d\theta \, d\phi. 
\label{dVMT}
\end{equation}
This volume element is an invariant quantity under any change of parametrization. 
Moreover, by virtue of the formula
\begin{equation}
\sqrt{\det{\left[ g_{\rm MT}{\left(  \bar{n}_1, \, \bar{n}_2, \, \theta, \, \phi \right)} \right] }}
=\frac{1}{16}\sqrt{H_{\bar{n}_1} H_{\bar{n}_2}H_{\theta}H_ {\phi}},
\label{JPMT}
\end{equation}
it is proportional to the square root of the determinant of the QFI matrix: 
\begin{align} 
& {\mathcal J}_{\rm MT}{\left(  \bar{n}_1, \bar{n}_2, \theta \right) }:
=\sqrt{H_{\bar{n}_1} H_{\bar{n}_2}H_{\theta}H_ {\phi}}:   \notag\\
&  {\mathcal J}_{\rm MT}{\left(  \bar{n}_1, \bar{n}_2, \theta \right) }
=\frac{1}{\sqrt{\bar{n}_1(\bar{n}_1+1)\bar{n}_2(\bar{n}_2+1)} }    \notag\\
& \times \frac{\left( \bar{n}_1-\bar{n}_2 \right)^2 \,\sin(\theta) }
{2\bar{n}_1 \bar{n}_2+ \bar{n}_1+\bar{n}_2 }.
\label{QJPMT}
\end{align}
The function\ (\ref{QJPMT}) is called quantum Jeffreys' prior \cite{Slater} due to its role 
in Bayesian statistical inference \cite{CFS2002}.  Indeed, by extension of Jeffreys' geometric 
rule \cite {Kass}, when properly normalized, it is a reliable {\em a priori} probability density  
on any compact part of the state manifold 
${\mathcal M} {\left(  \bar{n}_1, \, \bar{n}_2, \, \theta, \, \phi \right)}$.

\subsection{Squeezed thermal states}

The fidelity\ (\ref{BDn}) between neighbouring STSs, 
\begin{align}
& {\cal F}(t):={\cal F} \left[ \hat{\rho} \left(  \bar{n}_1, \, \bar{n}_2, \, r, \, \phi \right), \right.  \notag \\
& \left. \hat{\rho} \left(  \bar{n}_1+t d \bar{n}_1, \, \bar{n}_2+t d \bar{n}_2, \,    
r+t dr, \, \phi+t d\phi \right)  \right],     \notag \\
& \left( t \geqq 0 \right),
\label{FnSTS}
\end{align}
is given by the formula\ (\ref{F2(Kt)}),
where the functions $K_{\pm}(t)$ are consistent with Eq.\ (\ref{FnSTS}).
Accordingly, the Bures metric\ (\ref{dsB})  for STSs has the expression\ (\ref{dsSG}).
It is easy to recover the identities\ (\ref{F(0)}),  the first one being included in Eq.\ (\ref{FST=1}). 
We are subsequently lead to the formula: 
\begin{align}
& \left[ \left( ds_{\rm B} \right)_{\rm ST} \right]^2=\frac{1}{4}\left( \frac{1}{\bar{n}_1(\bar{n}_1+1)}
\left( d\bar{n}_1 \right)^2
+\frac{1}{\bar{n}_2(\bar{n}_2+1)}\left( d\bar{n}_2 \right)^2   \right.  \notag \\     
& \left. +\frac{\left( \bar{n}_1+\bar{n}_2+1 \right)^2 }
{2\bar{n}_1 \bar{n}_2+\bar{n}_1+\bar{n}_2+1} 
\left\{ [d(2r)]^2 +\left[ \sinh(2r) \right]^2  (d\phi)^2 \right\} \right).
\label{dsST}
\end{align}
Equation\ (\ref{dsST}) actually defines  the Bures metric on the Riemannian manifold
${\mathcal M} {\left(  \bar{n}_1, \, \bar{n}_2, \, 2r, \, \phi \right)}$ of the two-mode STSs. 
Note that the associated metric tensor has a diagonal matrix 
$g_{\rm ST}{\left(  \bar{n}_1, \, \bar{n}_2, \, 2r, \, \phi \right)}$. The sum of the first two terms
in the r. h. s. of Eq.\ (\ref{dsST}) is the squared line element\ (\ref{dsT}) on the manifold  
${\mathcal M} {\left(  \bar{n}_1, \, \bar{n}_2 \right) }$ of the two-mode TSs\ (\ref{TS}).
The interaction of the incident thermal modes with the non-degenerate parametric amplifier 
produces  the $SU(1,1)$ unitary state evolution\ (\ref{STS}). This is represented 
by the last two terms in the r. h. s. of Eq.\ (\ref{dsST}).  Remarkably, their sum is proportional 
to the Minkowski metric on the hyperboloid of two sheets $x^2+y^2-z^2=-1$:        
\begin{equation}
\left( ds_{\tau,  \phi} \right)^2= (d\tau)^2 \, +\left[ \sinh(\tau) \right]^2  (d\phi)^2:    \quad  \tau=2r.
\label{H2}
\end{equation}
The upper sheet $z>0$ of the hyperboloid is a two-dimensional Riemannian manifold 
denoted $H_{-1}^2$, which is a non-compact homogeneous space: $H_{-1}^2=SU(1,1)/U(1)$.
At the same time,  $H_{-1}^2$ is an analytic model of the hyperbolic plane $H^2$ 
\cite{Coxeter,Cannon}. 

By reason of the general relation\ (\ref{1:4}),  Eq.\ (\ref{dsST}) supplies the infinitesimal 
statistical distance
\begin{align}
& \left[ \left( ds_{\rm F} \right)_{\rm ST} \right]^2=H_{\bar{n}_1} \left( d\bar{n}_1 \right)^2
+H_{\bar{n}_2} \left( d\bar{n}_2 \right)^2    \notag \\     
& +H_{2r} \left[ d(2r) \right]^2 +H_{\phi} (d\phi)^2,
\label{dsFST}
\end{align}
whose QFI matrix is diagonal, with entries independent of the phase $\phi$: 
\begin{align}
& H_{\bar{n}_1}=\frac{1}{\bar{n}_1(\bar{n}_1+1)} \, ,     \qquad
H_{\bar{n}_2}=\frac{1}{\bar{n}_2 (\bar{n}_2+1)} \, ,       \notag \\     
& H_{2r}=\frac{\left( \bar{n}_1+\bar{n}_2 +1\right)^2 }
{2\bar{n}_1 \bar{n}_2+\bar{n}_1+\bar{n}_2+1}\, ,     \notag \\ 
& H_{\phi}=\frac{\left( \bar{n}_1+\bar{n}_2 +1\right)^2 \, \left[ \sinh(2r) \right]^2 }
{2\bar{n}_1 \bar{n}_2+\bar{n}_1+\bar{n}_2+1}.
\label{HST}
\end{align}
This diagonal form of the QFI tensor shows that the natural parameters 
$\left\{  \bar{n}_1, \, \bar{n}_2, \, 2r, \, \phi \right \} $ of the STSs are orthogonal.
It allows one to write directly the quantum Cram\'{e}r-Rao bound for the variance 
of any such a state estimator ${\xi}_{\alpha}$  \cite{Paris}:
\begin{equation}
\left( \Delta {\xi}_{\alpha} \right)^2 \geqq \frac{1}{{\cal N} H_{{\xi}_{\alpha}}},   \qquad
({\xi}_{\alpha}=\bar{n}_1, \, \bar{n}_2, \, 2r, \, \phi). 
\label{CRBS}
\end{equation}
In Eq.\ (\ref{CRBS}),  ${\cal N}$ denotes the number of the performed measurements.

To recapitulate, we point out that the Bures metric\ (\ref{dsST}) has a decomposition 
similar to that shown by Eq.\ (\ref{dsMT+}):
\begin{align} 
& \left[ \left( ds_{\rm B} \right)_{\rm ST} \right]^2=\left[ \left( ds_{\rm B} \right)_{\rm T} \right ]^2
+\left[ f_{\rm ST}{\left(  \bar{n}_1, \bar{n}_2 \right) } \right]^2 \left( ds_{2r,  \phi} \right)^2:   \notag \\
& f_{\rm ST}{\left(  \bar{n}_1, \bar{n}_2 \right) }:=\frac{1}{2}\sqrt{H_{2r}}.
\label{dsST+}
\end{align}
Besides, we indicate the Bures-metric volume element on the Riemannian manifold 
${\mathcal M} {\left(  \bar{n}_1, \, \bar{n}_2, \, 2r, \, \phi \right)}$ of the two-mode STSs:
\begin{equation}
d{\mathcal V}_{\rm B}:=\sqrt{\det{\left[ g_{\rm ST}
{\left(  \bar{n}_1, \bar{n}_2,  2r, \phi \right)} \right] }} \, d\bar{n}_1 \, d\bar{n}_2 \, d(2r) d\phi. 
\label{dVST}
\end{equation}
The formula
\begin{equation}
\sqrt{\det{\left[ g_{\rm ST}{\left(  \bar{n}_1, \, \bar{n}_2, \, 2r, \, \phi \right)} \right] }}
=\frac{1}{16}\sqrt{ H_{\bar{n}_1} H_{\bar{n}_2}H_{2r}H_ {\phi} }
\label{JPST}
\end{equation}   
demonstrates that the parametrization-invariant volume element\ (\ref{dVST}) 
is proportional to the quantum Jeffreys' prior:
\begin{align} 
& {\mathcal J}_{\rm ST}{\left(  \bar{n}_1, \bar{n}_2, 2r \right) }:
=\sqrt{ H_{\bar{n}_1} H_{\bar{n}_2} H_{2 r} H_ {\phi} }.
\label{QJPST}
\end{align}
We recall the separability threshold $r_s$ for a two-mode STS, introduced 
in Ref. \cite{PTH2001}:
\begin{equation}
\sinh(r_s):=\sqrt{ \frac{\bar{n}_1 \, \bar{n}_2}{ \bar{n}_1+\bar{n}_2+1} }.
\label{r_s}
\end{equation}
Noticeably, the quantum Jeffreys' prior\ (\ref{QJPST}) depends only on two variables, 
$r_s$ and $r$:
\begin{equation}
{\mathcal J}_{\rm ST}{\left( 2r_s, 2r \right) }=\frac{4\sinh(2r)}{\sinh(4r_s)}.
\label{JST}
\end{equation} 
At a fixed value of the squeeze parameter $r$, the function\ (\ref{JST}) strictly decreases 
with the variable $r_s$ from the limit ${\mathcal J}_{\rm ST}(0, 2r) =+\infty$  to zero, 
for $r_s \to +\infty$. The starting limit is reached by any two-mode STS at the physicality edge 
$(r_s=0)$ and, in particular, by any two-mode SVS. The value at the separability threshold
${\mathcal J}_{\rm ST}{\left( 2r_s, 2r_s \right) }=2\,{\rm sech}(2r_s)$ is itself a decreasing 
function of the variable $r_s$.

\subsection{Discussion}

We stress that the explicit formula\  (\ref{2F}) for the fidelity 
of two-mode GSs allows one the evaluation of QFI for estimating various parameters 
via Eq.\ (\ref{dsB}). This method has efficiently been exploited in some recent 
applications \cite{Adesso1,Fu1}. For instance, the concept of interferometric power, 
introduced and evaluated in Ref. \cite{Adesso1},  reduces in the particular case of an STS
to the QFI matrix element $H_{\phi}$, Eq.\  (\ref{HST}).  A productive research \cite{Fu2,Fu3}
in relativistic quantum metrology is based on QFI obtained by using Eq.\ (\ref{2F}).  

However, there are few cases when one could use an explicit expression 
of the Uhlmann fidelity to derive the QFI metric via Eq.\ (\ref{dsB}). The most widespread 
approach to evaluating the QFI is based on a central quantity in parameter estimation theory, 
namely, the symmetric logarithmic derivative (SLD) \cite{BC1994,Paris}.  
In particular, some important results have been obtained for GSs  by employing the SLD-method. 
An interesting example is the optimal estimation of entanglement for two-mode symmetric STSs 
in Ref. \cite{GGP}. The QFI for one-parameter estimation in the case of  multi-mode Gaussian 
channels and states was recently derived  \cite{MI,Mo,Jiang}.  The general result 
for an $n$-mode GS obtained in Refs. \cite{Mo,Jiang} is a compact expression in terms of the CM,
the displacement vector, and their first-order derivatives with respect to the estimated parameter.  

In order to check on Eqs.\ (\ref{HMT}) and\ (\ref{HST}), we apply the QFI formula 
from Ref. \cite{Jiang} together with all the necessary ingredients involved. Making use 
of the CMs\ (\ref{VMT}) and\ (\ref{VST}), as well as of the corresponding symplectic 
matrices\ (\ref{S(U)}) and\ (\ref{S(V)}) that diagonalize them by congruence, a routine calculation 
allows us to retrieve the QFI matrices for both manifolds of two-mode MTSs and STSs. 
However, the key point is the knowledge of the diagonalizing symplectic transformations. 

\section{Scalar curvatures of the Bures metric on the manifolds of special two-mode Gaussian states}

The scalar curvature is the simplest invariant derived from the metric of a Riemannian 
or pseudo-Riemannian manifold. This is a real function $R$ defined on such a manifold 
$\mathcal M$ which is determined {\em solely} by its intrinsic geometry. The value $R(p)$ 
at each point $p \in {\mathcal M}$ depends on the local features of the metric.

We evaluate the scalar curvatures on the four-dimensional Riemannian manifolds 
of the two-mode MTSs and STSs starting from their Bures metric tensors. 
Such a calculation exploits standard formulae from Riemannian Geometry \cite{Carmo} 
and consists of the following compulsory steps:
\begin{enumerate} 
\item Evaluation of the Christoffel symbols of the Levi-Civita connection;

\item Calculation of the required components of the Riemann curvature tensor 
by employing the Christoffel symbols and their first-order derivatives;

\item Calculation of the diagonal components of the Ricci tensor, which is defined 
as a contraction of the Riemann tensor;

\item Evaluation of the Riemannian scalar curvature as the trace of the Ricci tensor 
with respect to the metric.
\end{enumerate} 
We mention that calculations of the scalar curvature of the Bures metric tensor 
along the same lines were carried out previously for Riemannian manifolds 
of two kinds of single-mode GSs, namely, the STSs \cite{Twamley}  
and the displaced TSs \cite{PS1998}. 

\subsection{Mode-mixed thermal states} 

By carrying out the above-sketched program, we have found the scalar curvature 
on the  Riemannian manifold ${\mathcal M} {\left(  \bar{n}_1, \, \bar{n}_2, \, \theta, \, \phi \right)}$ 
of the two-mode MTSs: 

\begin{align}
& R_{\rm MT}{\left(  \bar{n}_1,  \bar{n}_2 \right)}=\frac{2}{\left( 2 \bar{n}_1 \bar{n}_2 
+ \bar{n}_1 +\bar{n}_2  \right)^2}    \notag \\
& \times \left[ \left(  \bar{n}_1-\bar{n}_2 \right)^2  -24\, \bar{n}_1\left( \bar{n}_1 +1 \right)
\bar{n}_2 \left( \bar{n}_2  +1 \right)  \right.     \notag \\
& \left.  +9\left( 2 \bar{n}_1 \bar{n}_2 + \bar{n}_1 +\bar{n}_2  \right)  \right].
\label{RMT}
\end{align}
The scalar curvature\ (\ref{RMT}) does not depend on the parameters $\{ \theta, \, \phi \}$
of the beam splitter. Its expression is valid for any values of the mean thermal photon 
occupancies $ \bar{n}_1,  \bar{n}_2$, and displays the symmetry property
\begin{equation}
 R_{\rm MT}{\left(  \bar{n}_2,  \bar{n}_1 \right)}= R_{\rm MT}{\left(  \bar{n}_1,  \bar{n}_2 \right)}.
\label{symRMT}
\end{equation}
Therefore, Eq.\ (\ref{RMT}) describes a two-dimensional surface in ${\mathbb R}^3$ 
which is represented in Fig. 1. It looks like a descending symmetric valley 
whose talweg is precisely the intersection with its symmetry plane $\bar{n}_1= \bar{n}_2$, 
i.e., the vertical plane that bisects the first octant.  
\begin{figure}[h]
\center
\includegraphics[width=7cm]{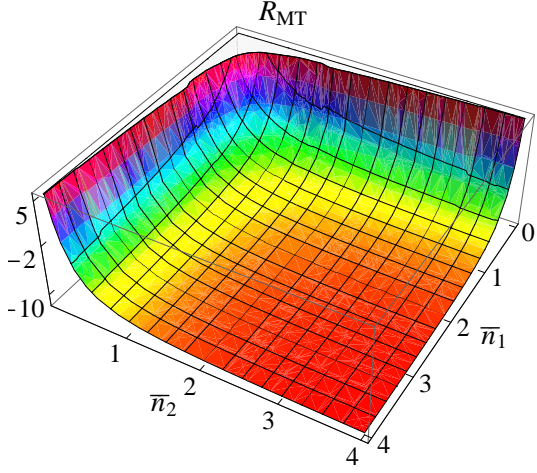}
\caption{(Color online) The scalar curvature $R_{\rm MT}{\left(  \bar{n}_1,  \bar{n}_2 \right)}$, 
Eq.\ (\ref{RMT}), of the two-mode MTSs. This is a convex surface whose general aspect 
is that of a symmetric valley that descends and widens continuously. 
Its talweg belongs to the symmetry plane $\bar{n}_1= \bar{n}_2$ and is drawn in Fig. 2a.}
\end{figure}

In the limit case $\bar{n}_1= \bar{n}_2=:\bar{n}$ of an emerging two-mode TS,  
Eq.\ (\ref{RMT}) simplifies to:
\begin{equation}
R_{\rm MT}{\left(  \bar{n},  \bar{n} \right)}= \frac{9}{\bar{n}\left( \bar{n}+1 \right) }-12.
\label{RT}
\end{equation}
The graph of the above function is the intersection of the two-dimensional 
surface\ (\ref{RMT}) and its symmetry plane $\bar{n}_1= \bar{n}_2$.
The  function\ (\ref{RT}) strictly decreases with the variable $\bar{n}$ from $+\infty$
at $\bar{n}=0$ to the negative asymptotic value 
$\lim_{\bar{n} \to \infty} R_{\rm MT}{\left( \bar{n}, \bar{n} \right) }=-12$. 
Besides, this is a convex function which has a unique zero,  $\bar{n}=\frac{1}{2}$. 

\begin{figure}[h]
\center
\includegraphics[width=4.4cm]{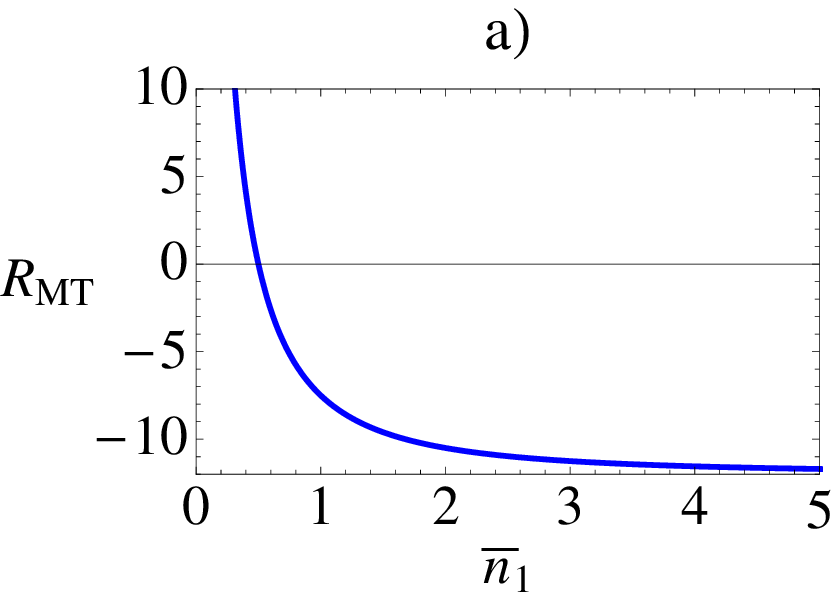}
\includegraphics[width=4cm]{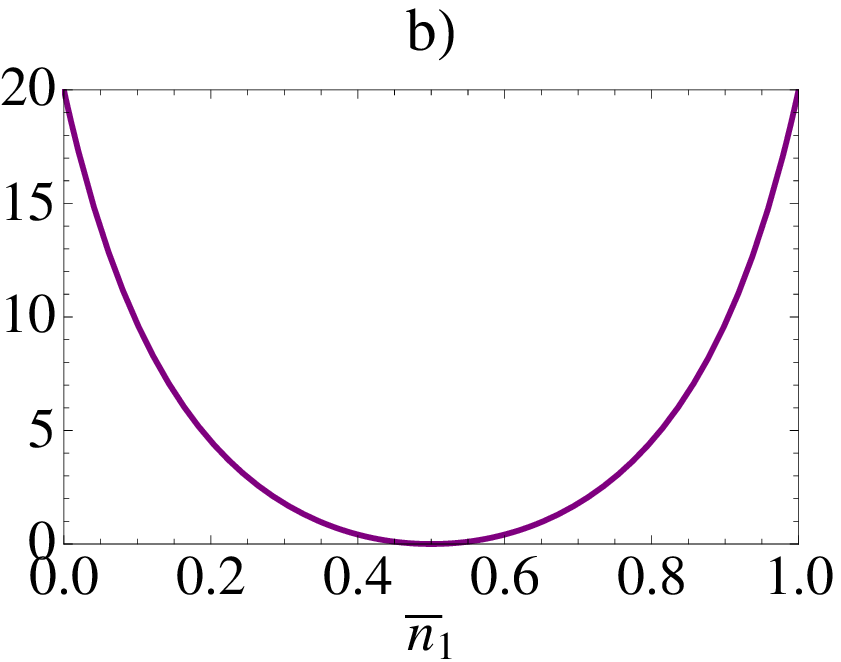}
\caption{(Color online) a) The vertical intersection\ (\ref{RT}) of the surface\ (\ref{RMT}) 
and its symmetry plane $\bar{n}_1= \bar{n}_2$. This is the talweg 
of the surface\ (\ref{RMT}) and  is made up of the symmetric two-mode TSs.\\
b) The vertical intersection\ (\ref{RMTperp}) of the surface\ (\ref{RMT}) 
and  the plane $\bar{n}_1+\bar{n}_2=1$. The vertical sections a) and b)
which are orthogonal and meet at their unique zero, $\bar{n}_1=\frac{1}{2}$.} 
\end{figure}

It is instructive to examine further the intersection of the surface\ (\ref{RMT}) 
and a vertical plane perpendicular to its symmetry plane. As an example, 
we choose the plane $\bar{n}_1+\bar{n}_2=1$ which meets the symmetry plane
in the vertical line $\bar{n}_1=\bar{n}_2=\frac{1}{2}$. This straight line contains
the above-mentioned zero of the function\ (\ref{RT}). The  intersection 
of the surface\ (\ref{RMT}) and the vertical plane $\bar{n}_1+\bar{n}_2=1$ 
is the graph of the following function of the variable  $\bar{n}_1 \in [0, \, 1]:$ 
\begin{align}
& R_{\rm MT}{\left(  \bar{n}_1, 1- \bar{n}_1 \right)}
=-4\, \frac{12 {\alpha}^2+17\alpha-5}{\left( 2\alpha+1 \right)^2} \geqq 0 ,        \notag \\  
& \alpha:=\bar{n}_1 \left( 1- \bar{n}_1 \right) \in \left[ 0, \, \frac{1}{4} \right].
\label{RMTperp}
\end{align}
The function\ (\ref{RMTperp}) strictly decreases from the limit value $R_{\rm MT}(0, \, 1)=20$
to its minimum $R_{\rm MT}{\left(  \frac{1}{2}, \, \frac{1}{2} \right) }=0$
and then has a mirror increase on the interval  $\bar{n}_1 \in \left[  \frac{1}{2}, \, 1 \right].$ 
Its graph exhibits the profile of a symmetric valley. Such a vertical section is typical 
for the  the two-dimensional surface\ (\ref{RMT}). 
The vertical sections\ (\ref{RT}) and\ (\ref{RMTperp}) are plotted in Figs. 2a and 2b.

A noteworhy limit situation arises when one incoming mode is in the vacuum state
and the other is not $\left(  \bar{n}_1>0, \;  \bar{n}_2=0 \right)$. Then the output 
two-mode MTS is at the physicality edge and has the scalar curvature
\begin{equation}
R_{\rm MT}{\left(  \bar{n}_1, 0 \right)}=2+ \frac{18}{\; \bar{n}_1}.
\label{RMTedge}
\end{equation}
The function\ (\ref{RMTedge}) is positive, strictly decreasing and convex.
Figure 5 presents its graph, as well as that of the function\ (\ref{RSTedge}).

\subsection{Squeezed thermal states}

In a similar way we have evaluated the scalar curvature 
on the  Riemannian manifold ${\mathcal M} {\left(  \bar{n}_1, \, \bar{n}_2, \, 2r, \, \phi \right)}$ 
of the two-mode STSs: 

\begin{align}
& R_{\rm ST}{\left(  \bar{n}_1,  \bar{n}_2 \right)}=\frac{2}{\left( 2 \bar{n}_1 \bar{n}_2 
+ \bar{n}_1 +\bar{n}_2 +1 \right)^2}    \notag \\
& \times \left[ \left(  \bar{n}_1+\bar{n}_2 +1\right)^2  -24\, \bar{n}_1\left( \bar{n}_1 +1 \right)
\bar{n}_2 \left( \bar{n}_2  +1 \right)  \right.     \notag \\
& \left.  -9\left( 2 \bar{n}_1 \bar{n}_2 + \bar{n}_1 +\bar{n}_2+1  \right)  \right].
\label{RST}
\end{align}
The scalar curvature\ (\ref{RST}) does not depend on the parameters $\{ 2r, \, \phi \}$
of the  non-degenerate parametric amplifier. Its expression is valid for any values 
of the mean thermal photon occupancies $ \bar{n}_1,  \bar{n}_2$, and displays 
the symmetry property
\begin{equation}
 R_{\rm ST}{\left(  \bar{n}_2,  \bar{n}_1 \right)}= R_{\rm ST}{\left(  \bar{n}_1,  \bar{n}_2 \right)}.
\label{symRST}
\end{equation}
Accordingly, the two-dimensional surface\ (\ref{RST}) in ${\mathbb R}^3$ has 
the vertical symmetry plane $\bar{n}_1= \bar{n}_2$ that bisects the first octant.  
Figure 3 displays the general aspect of this surface. 
It looks like a pair of opposite symmetric valleys.  The ascending valley is narrow and steep,
while the descending one is broader and slower. The talweg is the intersection\ (\ref{RsST}) 
with the symmetry plane $\bar{n}_1= \bar{n}_2$, 
while the watershed is the intersection\ (\ref{RSTperp}) with the vertical plane 
$\bar{n}_1+\bar{n}_2=2\bar{n}_s$, which is perpendicular to the first one. 
The talweg and the watershed are tangent at the unique saddle point $S$ 
on the surface, which is located at the point $(\bar{n}_s, \bar{n}_s)$ 
with $\bar{n}_s:=-0.5+\sqrt{1.15} \approxeq 0.5724$. 

\begin{figure}[h]
\center
\includegraphics[width=8cm]{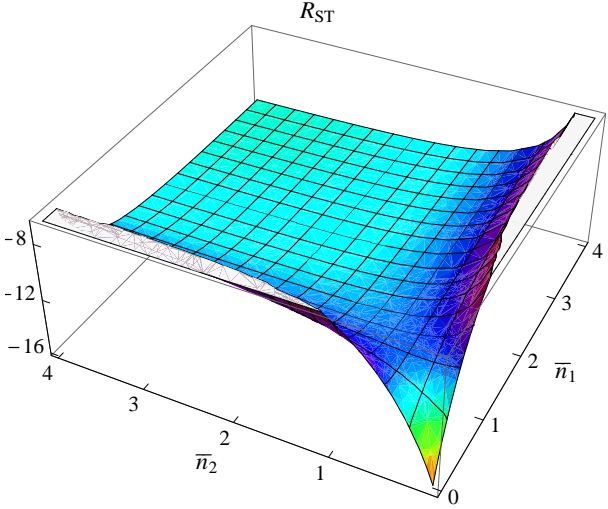}
\caption{(Color online) The scalar curvature $R_{\rm ST}{\left(  \bar{n}_1,  \bar{n}_2 \right)}$, 
Eq.\ (\ref{RST}), of the two-mode STSs. This surface consists of two opposite 
symmetric valleys separated by a watershed: a narrow, steeply ascending valley
and a wider one which is slowly descending. 
It possesses a unique saddle point $S$ at the intersection of the talweg 
and the watershed. The talweg is the normal section\ (\ref{RsST}) 
in the symmetry plane $\bar{n}_1= \bar{n}_2$, while the watershed 
is the normal section\ (\ref{RSTperp}) in the plane $\bar{n}_1+\bar{n}_2=2\bar{n}_s$. 
These normal sections, which are vertical and orthogonal, are represented
in Figs. 4a and 4b.}
\end{figure}

For symmetric two-mode STSs,  $\left( \bar{n}_1= \bar{n}_2=:\bar{n} \right)$,  
the scalar curvature\ (\ref{RST}) is negative:
\begin{align}
& R_{\rm ST}{\left(  \bar{n},  \bar{n} \right) }
=-4\, \frac{12 {\beta}^2+7\beta+4}{\left( 2\beta+1 \right)^2} < 0 ,      \notag \\  
& \beta:=\bar{n} \left( \bar{n}+1 \right)  \geqq 0.
\label{RsST}
\end{align}
The graph of the above function is reproduced in Fig. 4a and  is the talweg
of the surface\ (\ref{RST}). On its ascending side, the function\ (\ref{RsST})  
has a steep rise from the limit value $ R_{\rm ST}(0, \, 0)=-16$, 
reached for any two-mode SVS, to a maximum 
$R_{\rm ST}{\left(  \bar{n}_s,  \bar{n}_s \right) }=-\frac{143}{14} \approxeq -10.2143,$
reached at the point $\bar{n}_s=-0.5+\sqrt{1.15} \approxeq 0.5724$. 
Then, on the descending side, it has a moderate fall toward the asymptotic value 
$\lim_{\bar{n} \to \infty} R_{\rm ST}{\left(  \bar{n},  \bar{n} \right) }=-12$.
As expected on intuitive grounds, this asymptotic limit coincides with the similar one 
for MTSs, displayed in Fig. 2a:
$\lim_{\bar{n} \to \infty} R_{\rm MT}{\left(  \bar{n},  \bar{n} \right) }=-12$.

Let us contemplate next the intersection of the surface\ (\ref{RST}) 
and the vertical plane $\bar{n}_1+\bar{n}_2=2\bar{n}_s$, where $\bar{n}_s$
is the maximum point of the function\ (\ref{RsST}). This plane is perpendicular
to the symmetry plane $\bar{n}_1= \bar{n}_2$ and meets it in the vertical 
straight line $\bar{n}_1=\bar{n}_2=\bar{n}_s$. The aforementioned intersection 
is the graph of the following function of the variable  
$\bar{n}_1 \in \left[ 0, \, 2\bar{n}_s \right]:$ 
\begin{align}
& R_{\rm ST}{\left(  \bar{n}_1, 2\bar{n}_s- \bar{n}_1 \right)}
=-\frac{4}{\left[ 2\left( \gamma +\bar{n}_s \right) +1 \right]^2 }     \notag \\  
& \times  \left[ 12\left( \gamma +\bar{n}_s \right)^2 
+21\left( \gamma +\bar{n}_s \right) -8.6 \right] < 0\, ,    \notag \\  
& \gamma:=\bar{n}_1\left( 2\bar{n}_s- \bar{n}_1 \right) 
\in \left[ 0, \, {\left( \bar{n}_s \right) }^2 \right],    \quad       
 {\left( \bar{n}_s \right) }^2=0.3276 ,                 \notag \\ 
& 4-14\bar{n}_s\left( \bar{n}_s+1 \right) =8.6.
\label{RSTperp}
\end{align}
The function\ (\ref{RSTperp}) decreases from the limit value 
$R_{\rm ST}(0, 2\bar{n}_s) \approxeq -6.3925$ to its minimum, 
$R_{\rm ST}{\left( \bar{n}_s,  \bar{n}_s \right) }=-\frac{143}{14} \approxeq -10.2143,$
and then has a mirror increase on the interval  
$\bar{n}_1 \in \left[ \bar{n}_s, \, 2\bar{n}_s \right].$ 
Its graph is therefore the profile of a symmetric valley and is drawn in Fig. 4b.  

\begin{figure}
\center
\includegraphics[width=4.4cm]{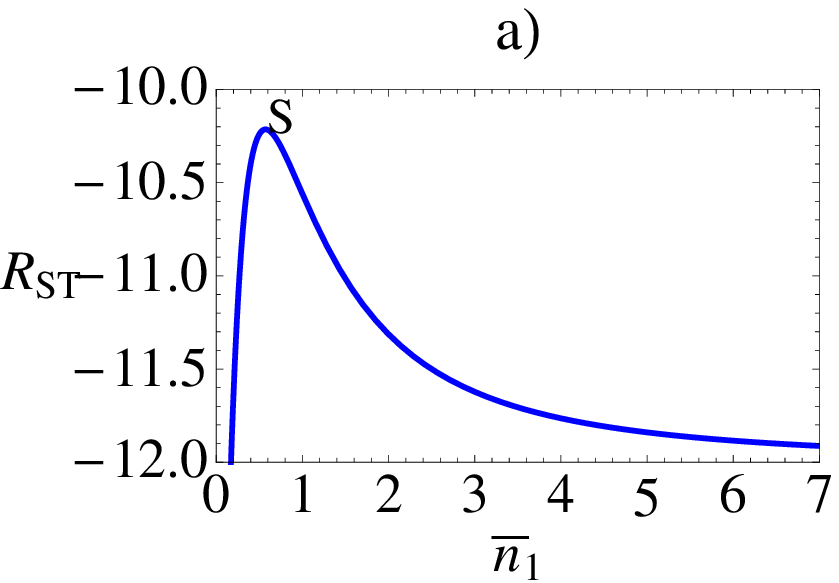}
\includegraphics[width=4cm]{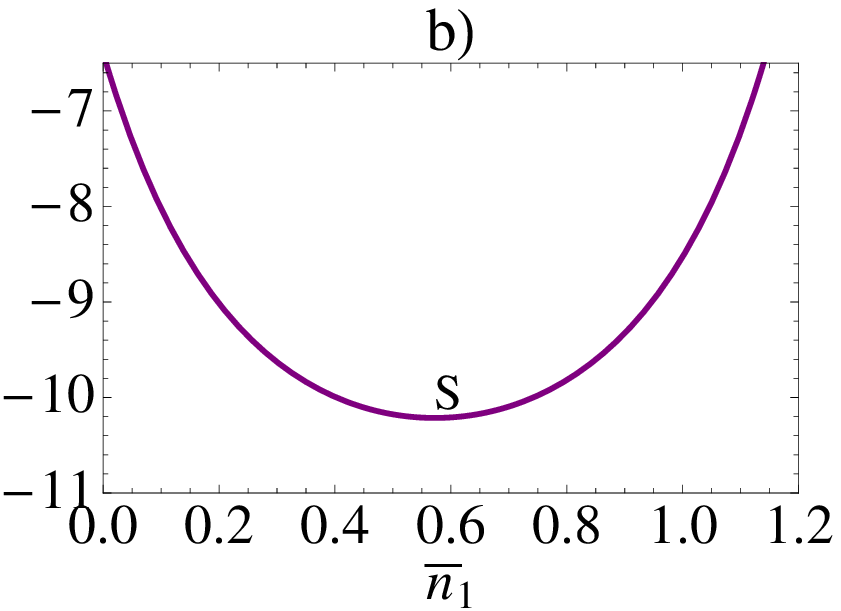}
\caption{(Color online) a) The vertical intersection\ (\ref{RsST}) of the surface\ (\ref{RST}) 
and its symmetry plane $\bar{n}_1= \bar{n}_2$. This is the talweg 
of the surface\ (\ref{RST}) and  is made up of the symmetric two-mode STSs.\\
The function\ (\ref{RsST})  has a steep rise from its lowest value 
$R_{\rm ST}(0, \, 0)=-16$, reached for any two-mode SVS, to a maximum 
$R_{\rm ST}{\left(  \bar{n}_s,  \bar{n}_s \right) }=-\frac{143}{14} \approxeq -10.2143,$
reached at the point $\bar{n}_s=-0.5+\sqrt{1.15} \approxeq 0.5724$. 
Then it has a moderate fall toward the asymptotic value 
$\lim_{\bar{n} \to \infty} R_{\rm ST}{\left(  \bar{n},  \bar{n} \right) }=-12$.
The function\ (\ref{RsST}) changes concavity at the unique inflection point 
$\bar{n}_i  \approxeq 0.9565$, where it has the value 
$R_{\rm ST}{\left(  \bar{n}_i,  \bar{n}_i \right) } \approxeq -10.5140$.\\
b) The vertical intersection\ (\ref{RSTperp}) of the surface\ (\ref{RST}) 
and  the plane $\bar{n}_1+\bar{n}_2=2\bar{n}_s$. This is the watershed
on the surface\ (\ref{RST}),  which is orthogonal to the talweg and touches it
at their common extremum point, $\bar{n}_1=\bar{n}_s$.}
\end{figure} 

Although such a vertical section is typical for the two-dimensional surface\ (\ref{RST}), 
the section\ (\ref{RSTperp}) is a rather special one. Indeed, the common extremum point  
$S: \left\{ \bar{n}_s, \, \bar{n}_s, \, R_{\rm ST}{\left( \bar{n}_s,  \bar{n}_s \right) } \right\} $
of the curves\ (\ref{RsST}) and\ (\ref{RSTperp}) is a saddle point on the two-dimensional 
surface\ (\ref{RST}) in ${\mathbb R}^3$, having an upward vertical normal. 
Moreover, this turns out to be its unique stationary point.  
The curves\ (\ref{RsST}) and\ (\ref{RSTperp}) are precisely the normal sections 
that include the principal directions  tangent to the surface\ (\ref{RST}) 
at the saddle point $S$ \cite{K}. 

When one of the incoming modes is in the vacuum state and the other is not, 
$\left(  \bar{n}_1>0, \;  \bar{n}_2=0 \right)$,  then the output two-mode STS 
is at the physicality edge. It  has the scalar curvature
\begin{equation}
R_{\rm ST}{\left(  \bar{n}_1, 0 \right)}=2- \frac{18}{ \bar{n}_1+1}.
\label{RSTedge}
\end{equation}
The function\ (\ref{RSTedge}) strictly increases with the variable $\bar{n}_1$ 
from the SVS value $R_{\rm ST}(0, \, 0)=-16$  to the positive asymptotic value 
$\lim_{\bar{n}_1 \to \infty} R_{\rm ST}{\left(  \bar{n}_1, 0 \right) }=2$.
Besides, it is a concave function which has a unique zero,  $\bar{n}_1=8$.
The curves\ (\ref{RMTedge}) and\ (\ref{RSTedge}) have a common asymptote
and are represented together in Fig. 5. 

\begin{figure}
\center
\includegraphics[width=8cm]{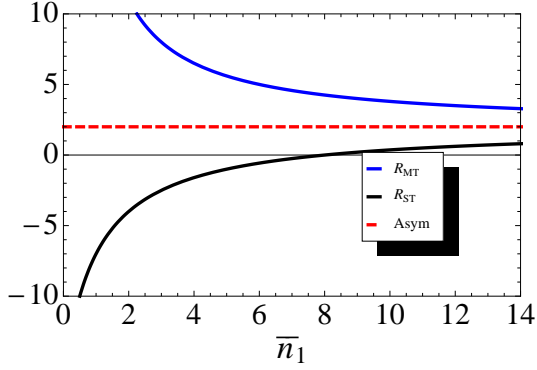}
\caption{(Color online) The scalar curvatures\ (\ref{RMTedge}) and\ (\ref{RSTedge}) 
of the two-mode MTSs and STSs at the physicality edge. They are figured
as the marginal vertical curves on the surfaces\ (\ref{RMT}) and\ (\ref{RST}), 
respectively, which meet the plane $ \bar{n}_2=0$ along them. 
The former is decreasing and convex, while the latter is increasing and concave.
Their starting values are, respectively, $R_{\rm MT}(0,0) = +\infty$ 
for the vacuum state and $R_{\rm ST}(0,0) = -16$ for any two-mode SVS.
Nevertheless, they have a common asymptotic limit: 
$R_{\rm MT}(+\infty,0)=R_{\rm ST}(+\infty,0)=2.$} 
\end{figure}

\subsection{Alternative derivation}

The most intriguing feature of both scalar curvatures 
$R_{\rm MT}{\left(  \bar{n}_1,  \bar{n}_2 \right) }$, Eq.\ (\ref{RMT}),
and $R_{\rm ST}{\left(  \bar{n}_1,  \bar{n}_2 \right) }$, Eq.\ (\ref{RST}),
is that they do not depend on the parameters $\{ \theta, \, \phi \}$ and, 
respectively, $\{ 2r, \, \phi \}$. These parameters specify the unitary transformations modeling 
the interactions of the incident thermal-light beams with the optical devices 
under discussion. Nevertheless, the scalar curvatures originate precisely 
in the mentioned unitary transformations. 
Since the scalar curvature of a given metric on a Riemannian manifold
is determined solely by the metric itself, any explanation should start from
the decompositions\ (\ref{dsMT+}) and\ (\ref{dsST+}) of the squared
Bures line elements on the Riemannian  manifolds of special two-mode GSs.
They can be written in terms of the metric tensors as follows:
\begin{align}
& g_{\rm MT}{\left(  \bar{n}_1, \, \bar{n}_2, \, \theta, \, \phi \right) }    \notag \\  
& = g_{\rm T}{\left(  \bar{n}_1, \, \bar{n}_2 \right) } \oplus  
\left[ f_{\rm MT}{\left(  \bar{n}_1, \bar{n}_2 \right) } \right]^2 \, 
g_{S^2} {\left( \theta, \, \phi \right)};
\label{gMT+}
\end{align}
\begin{align}
& g_{\rm ST}{\left(  \bar{n}_1, \, \bar{n}_2, \, 2r, \, \phi \right) }    \notag \\  
& = g_{\rm T}{\left(  \bar{n}_1, \, \bar{n}_2 \right) } \oplus  
\left[ f_{\rm ST}{\left(  \bar{n}_1, \bar{n}_2 \right) } \right]^2 \, 
g_{H_{-1}^2} {\left( 2r, \, \phi \right) }.
\label{gST+}
\end{align}
In the above equations, $g_{S^2} $ and $g_{H_{-1}^2}$ designate the metric tensors
on the two-dimensional Riemannian manifolds $S^2$ and $H_{-1}^2$, as given 
by Eqs.\ (\ref{S2}) and\ (\ref{H2}),  respectively. In contrast to the Euclidean plane, 
these surfaces have a constant scalar curvature which is not zero. Moreover,
up to a scale factor, they are the only connected ones to share this intrinsic geometric property:
\begin{equation}
R{\left( {\mathbb R}^2 \right) }=0, \quad R{\left( S^2 \right) }=2, \quad 
R{\left( H_{-1}^2 \right) }=-2. 
\label{RSH}
\end{equation}

We further need the notion of warped product of two Riemannian manifolds,
which was introduced in Ref. \cite{BO'N}. Let $({\mathcal B}, \, g_{\rm B})$
and $({\mathcal F},\,  g_{\rm F})$  be Riemannian manifolds 
of dimensions $m$ and $n$, respectively, and 
$f:{\mathcal B} \to {\mathbb R}_{+}\setminus \{0\}$ a smooth function:  
$f \in C^{\infty}({\mathcal B})$. The warped product 
${\mathcal M}:={\mathcal B} {\times}_f {\mathcal F}$ is the differentiable 
product manifold ${\mathcal B} \times {\mathcal F}$, of dimension $m+n$, 
endowed with the Riemannian metric $g_{\rm M}:=g_{\rm B} \oplus f^2 g_{\rm F}$.
By means of this rule, the warping function $f>0$ determines 
the Riemannian structure of the warped product $({\mathcal M},\,  g_{\rm M})$.

The relationship between the scalar curvatures $R({\mathcal B}), \, R({\mathcal F})$, 
and $R({\mathcal M})$ was established in Ref. \cite{DD}: 
\begin{align}
& -\frac{4n}{n+1}{\Delta}_{g_{\rm B}}u+R({\mathcal B})u+R({\mathcal F})u^{\frac{n-3}{n+1}}
=R({\mathcal M})u,      \notag\\  
& u:=f^{\frac{n+1}{2}}.
\label{R(M)}
\end{align}
In Eq.\ (\ref{R(M)}), ${\Delta}_{g_{\rm B}}$ is the Laplace-Beltrami operator
on the Riemannian manifold $({\mathcal B}, \, g_{\rm B})$. This second-order 
differential operator is the divergence of the gradient:
\begin{align}
& {\Delta}_{g_{\rm B}}v =\frac{1}{\sqrt{\det{\left( g_{\rm B}\right) } } }
{\partial}_j \left[ \sqrt{\det{\left( g_{\rm B}\right) } } \left( g_{\rm B}\right)^{jk} 
{\partial}_k v \right],     \notag\\  
& v \in C^2({\mathcal B}).
\label{LB}
\end{align}

We mention two important consequences of Eq.\ (\ref{R(M)}). 

First, when $f=1$, it reduces to the familiar addition law
\begin{equation}
R{\left( {\mathcal B}\times {\mathcal F} \right) }=R{\left( {\mathcal B} \right) }
+R{\left( {\mathcal F} \right) }
\label{f=1}
\end{equation}
for the scalar curvature of the Riemannian product manifold 
$\left( {\mathcal B}\times {\mathcal F}, \, g_{\rm B} \oplus  g_{\rm F} \right)$.

Second, the scalar curvature  $R({\mathcal M})$ of the warped product
${\mathcal M}:={\mathcal B} {\times}_f {\mathcal F}$
does not depend on the parameters of the manifold ${\mathcal F}$
if and only if  ${\mathcal F}$ has a constant scalar curvature.  

Coming back to our current problem, the structure of the metric 
tensors\ (\ref{gMT+}) and\ (\ref{gST+}) shows that 
the four-dimensional Riemannian manifolds
${\mathcal M} {\left(  \bar{n}_1, \, \bar{n}_2, \, \theta, \, \phi \right)}$ and
${\mathcal M} {\left(  \bar{n}_1, \, \bar{n}_2, \, 2r, \, \phi \right)}$ 
are warped products:
\begin{equation}
{\mathcal M} {\left(  \bar{n}_1, \, \bar{n}_2, \, \theta, \, \phi \right)}
={\mathcal M} {\left(  \bar{n}_1, \, \bar{n}_2  \right) } 
{\times}_{ f_{\rm MT} } \, S^2( \theta, \, \phi);                                                                                                                            
\label{wMTS}
\end{equation}
\begin{equation}
{\mathcal M} {\left(  \bar{n}_1, \, \bar{n}_2, \, 2r, \, \phi \right)}
={\mathcal M} {\left(  \bar{n}_1, \, \bar{n}_2  \right) } 
{\times}_{ f_{\rm ST} } \,  H_{-1}^2(2r, \, \phi).
\label{wSTS}
\end{equation}
In view of the isometry\ (\ref{dsTR^2}), the two-dimensional Riemannian manifold
${\mathcal M} {\left(  \bar{n}_1, \, \bar{n}_2  \right) } \equiv {\mathbb R}_{+}^2$
of the two-mode TSs\ (\ref{TS}) has a vanishing scalar curvature.
However, the essential issue is that both $S^2$ and $H_{-1}^2$ are surfaces 
of constant scalar curvature. Then the second consequence stated above 
explains why the scalar curvatures 
$R_{\rm MT}{\left(  \bar{n}_1,  \bar{n}_2 \right) }$ 
and $R_{\rm ST}{\left(  \bar{n}_1,  \bar{n}_2 \right) }$ depend only on the
mean thermal photon numbers in the incoming modes. Needless to say, 
this dependence is specific for each of the warped products\ (\ref{wMTS}) 
and\ (\ref{wSTS}).

Let us check on our formulae\ (\ref{RMT}) and\ (\ref{RST}) by specializing 
Eqs.\ (\ref{R(M)}) and\ (\ref{LB})  for the warped products\ (\ref{wMTS}) and\ (\ref{wSTS}), 
respectively. Taking account of Eqs.\ (\ref{HMT}), \ (\ref{dsMT+}), \ (\ref{HST}), 
\ (\ref{dsST+}), and\ (\ref{RSH}), we find the following pair of equations 
for the scalar curvatures:
\begin{align}
& R_{\rm MT}{\left(  \bar{n}_1,  \bar{n}_2 \right) }=\frac{8}{H_{\theta}}
-2\sum_{j=1}^2 \bar{n}_j\left(  \bar{n}_j +1 \right) \left\{ 4\frac{ {\partial}^2}{\partial  \bar{n}_j^2}
\ln{\left( H_{\theta} \right)}   \right.   \notag\\  
& \left. +3\left[ \frac{\partial}{\partial  \bar{n}_j}\ln{\left( H_{\theta} \right)}  \right]^2 \right\} 
-4\sum_{j=1}^2 \left( 2\bar{n}_j +1 \right) \frac{\partial}{\partial \bar{n}_j}
\ln{\left( H_{\theta} \right)},      \notag\\   
& \left( \bar{n}_1 \ne \bar{n}_2 \right);
\label{eqRMT}
\end{align}
\begin{align}
& R_{\rm ST}{\left(  \bar{n}_1,  \bar{n}_2 \right) }=-\frac{8}{H_{2r} }
-2\sum_{j=1}^2 \bar{n}_j\left(  \bar{n}_j +1 \right) \left\{ 4\frac{ {\partial}^2}{\partial  \bar{n}_j^2}
\ln{\left( H_{2r} \right)}   \right.   \notag\\  
& \left. +3\left[ \frac{\partial}{\partial  \bar{n}_j}\ln{\left( H_{2r} \right)}  \right]^2 \right\} 
-4\sum_{j=1}^2 \left( 2\bar{n}_j +1 \right) \frac{\partial}{\partial \bar{n}_j}
\ln{\left( H_{2r} \right)}. 
\label{eqRST}
\end{align}
Substitution of the functions $H_{\theta}$ and $H_{2r}$ into the above equations
yields the expected  formulae\ (\ref{RMT}) and\ (\ref{RST}).

\section{Summary and conclusions}

We start this overview by stressing the main results we have obtained
in the present work. First, we have established an alternative expression 
of the fidelity between two-mode GSs, Eq.\ (\ref{2F(K)}). On the one hand,
this is efficient in evaluating the fidelity between special states,  
as it happens with the two-mode MTSs, Eqs.\ (\ref{K(+)M})-\ (\ref{K(-)M}), 
and the two-mode STSs, Eqs.\ (\ref{K(+)S})-\ (\ref{K(-)S}). 
On the other hand, it is flexible enough to have checked with ease, 
in Appendix C,  the inequality
${\cal F}({\hat{\rho}}^{\prime}, {\hat{\rho}}^{\prime \prime}) \leqq  1$
for both families of special two-mode GSs.

Second, taking advantage of the above-cited formulae,  we have derived 
the Bures infinitesimal geodesic distances on the Riemannian manifolds 
 ${\mathcal M} {\left(  \bar{n}_1, \, \bar{n}_2, \, \theta, \, \phi \right)}$ 
of the two-mode MTSs, Eq.\ (\ref{dsMT}), and 
${\mathcal M} {\left(  \bar{n}_1, \, \bar{n}_2, \, 2r, \, \phi \right)}$ 
of the two-mode STSs,  Eq.\ (\ref{dsST}). They are statistically relevant
due to the proportionality between the Bures and QFI metric 
tensors \cite{BC1994}. This endows the Bures metric with the general feature 
of statistical distinguishability between neigbouring states 
on a Riemannian manifold when performing suitable quantum 
measurements. In addition, the diagonal form of the QFI metric 
tensors\ (\ref{HMT}) of MTSs and\ (\ref{HST}) of STSs  with respect
to their natural parameters simplifies the corresponding 
quantum Cram\'{e}r-Rao inequalities.

Third, we have employed a standard procedure to evaluate the scalar
curvature associated to the Bures metric on each of the  Riemannian manifolds 
 ${\mathcal M} {\left(  \bar{n}_1, \, \bar{n}_2, \, \theta, \, \phi \right)}$ 
of the two-mode MTSs  and 
${\mathcal M} {\left(  \bar{n}_1, \, \bar{n}_2, \, 2r, \, \phi \right)}$ 
of the two-mode STSs. The formulae\ (\ref{RMT}) and\ (\ref{RST}) are 
the corresponding exact analytic results. Both scalar curvatures are 
merely functions of the mean photon numbers in the incident thermal modes, 
$\bar{n}_1$ and  $\bar{n}_2$. In spite of being determined by the interaction 
of thermal radiation with the optical instruments described previously,  neither 
of them depends on the specific parameters of the optical device involved.  
This particular property stems from the symmetry nature of the unitary 
operators\ (\ref{M_12}) and\ (\ref{S_12}) describing the optical processes 
in question:  $SU(2)$ and, respectively, $SU(1,1)$. In addition, we have exploited 
these symmetries to recover the scalar curvatures\ (\ref{RMT}) and\ (\ref{RST})
by an alternative method. Figures 1 and 3 allow one to visualize 
each of them as a function of the mean photon occupancies 
of the incoming thermal modes.

In order to reveal the significance of the Bures scalar curvature,
we follow closely Petz's exposition in Ref. \cite{Petz}. 
Let us consider an $n$-dimensional Riemannian manifold $({\mathcal M}, g_{\rm B})$ 
of quantum states, which is equipped with the Bures metric $g_{\rm B}$. 
 Then the geodesic distance $D_{\rm B}(\hat{\rho}^{\prime}, \hat{\rho}^{\prime \prime})$ 
between two points on the manifold ${\mathcal M}$ is interpreted 
as the statistical distinguishability of the two states by means of the optimal 
quantum measurement.

We focus on a given state $\hat{\rho}_0 \in ({\mathcal M}, \, g_{\rm B})$. The geodesic ball
\begin{equation}
B_n{ \left( \hat{\rho}_0; \, \varepsilon \right) }:=\{\hat{\rho} \in ({\mathcal M}, \, g_{\rm B}):  
\;\;  D_{\rm B}(\hat{\rho}_0, \hat{\rho})<\varepsilon\}
\label{B_n}
\end{equation}
contains all the states that can be distinguished from $\hat{\rho}_0$ 
by an information effort smaller than that corresponding 
to the radius $\varepsilon >0$. According to Jeffreys' rule \cite {Kass},
the size of the statistical inference region\ (\ref{B_n}), which measures 
the uncertainty in the information acquired about the state $\hat{\rho}_0$, 
is precisely the Bures volume 
$ {\mathcal V}_{\rm B}{\left[  B_n{ \left( \hat{\rho}_0; \, \varepsilon \right) }\right] }$.
Therefore, this volume can be interpreted as the {\em average statistical uncertainty}
of the state $\hat{\rho}_0 \in ({\mathcal M}, \,g_{\rm B})$. In order to improve 
the accuracy in identifying the state $\hat{\rho}_0$, one has to contract 
the geodesic ball\ (\ref{B_n}). Ideal asymptotic inference means reducing its radius 
as much as possible, that is, making $ \varepsilon \to 0$. This asymptotic behaviour 
is described by the following geometric formula \cite {GHL}: 
\begin{align} 
& {\mathcal V}_{\rm B}{ \left[ B_n{ \left( \hat{\rho}_0; \, \varepsilon \right) }\right] }
= V_{n}(1)\,  {\varepsilon}^n
-\frac{ V_{n}(1) }{n+2} \, R{ \left( \hat{\rho}_0 \right) }\, {\varepsilon}^{n+2} 
+{\rm o}{ \left( {\varepsilon}^{n+2} \right) }:     \notag \\
&  V_{n}(1)=\frac{ {\pi}^{ \frac{n}{2} } }{\Gamma{ \left( \frac{n}{2}+1 \right) } },                       
\qquad       \left( \varepsilon \ll 1 \right).
\label{VB_n}
\end{align}
In Eq.\ (\ref{VB_n}), $V_{n}(1)$ is the volume of the unit ball $B_n(0;1)$
in the Euclidean space  ${\mathbb R}^n$, while $R{ \left( \hat{\rho}_0 \right) }$ 
denotes the Bures scalar curvature at the state $\hat{\rho}_0$.
What Eq.\ (\ref{VB_n}) shows us is that, under the condition $\varepsilon \ll 1$, 
the average statistical uncertainty  is fully determined by the scalar curvature  
$R{ \left( \hat{\rho}_0 \right) }$ and, namely, is a decreasing  function of it.

To conclude, we come back to our four-dimensional Riemannian manifolds of special
two-mode GSs, ${\mathcal M} {\left(  \bar{n}_1, \, \bar{n}_2, \, \theta, \, \phi \right)}$
and ${\mathcal M} {\left(  \bar{n}_1, \, \bar{n}_2, \, 2r, \, \phi \right)}$. 
Figures 1 and 3, where their scalar curvatures are plotted, offer a global view 
on the average statistical uncertainty of these noteworthy states. 
Remark that almost all of them are noisy, except for the pure states, 
$\left( \bar{n}_1=0, \, \bar{n}_2=0 \right) $,  i. e., the vacuum and, respectively,
all the two-mode SVSs. The figured values, albeit not completely intuitive, 
nevertheless display several regularities and provide some interesting comparisons. 
It is our opinion that these results urge a deeper understanding 
from a quantum information perspective.

\appendix 
\section{Fidelity between $n$-mode Gaussian states:
The inequality ${\cal F}({\hat{\rho}}^{\prime}, {\hat{\rho}}^{\prime \prime}) \geqq  
{\rm Tr}({\hat{\rho}}^{\prime} {\hat{\rho}}^{\prime \prime} )$}

We make a digression intended for a pair of arbitrary  $n$-mode Gaussian states,
$\hat{\rho}^{\prime}$ and $\hat{\rho}^{\prime \prime}$. It is straightforward to extend 
Eqs.\ (\ref{Delta})-\ (\ref{overlap}) to the multi-mode case  \cite{PT2012}. Since the overlap                                                      
${\rm Tr}({\hat{\rho}}^{\prime} {\hat{\rho}}^{\prime \prime} )$ of two GSs  never vanishes, 
we have been lead to introduce a key GS  \cite{PT2012}, 

\begin{align}
\hat{\rho}_B:=[{\rm Tr}(\hat{\mathcal B})]^{-1}\hat{\mathcal B}, \qquad 
\hat{\mathcal B}:=\sqrt{\hat{\rho}^{\prime \prime}}\hat{\rho}^{\prime}
\sqrt{\hat{\rho}^{\prime \prime}}.
\label{rhoB}
\end{align}
Its CM, denoted ${\mathcal V}_B$, has the symplectic invariants \cite{PT2012}:  

\begin{equation}
\det \left({\mathcal V}_B \right)=2^{-2n}\frac{\Gamma}{\Delta}, \quad 
\det \left({\mathcal V}_B +\frac{i}{2}J \right)=2^{-2n}\frac{\Lambda}{\Delta}.
\label{invar}
\end{equation}
Obviously, the fidelity\ (\ref{F}) of two GSs  is proportional to their overlap given by an $n$-mode analogue of Eq.\ (\ref{overlap}) \cite{PT2012}:
\begin{align}
{\cal F}(\hat{\rho}^{\prime}, \hat{\rho}^{\prime \prime})
=\left[ {\rm Tr}\left( \sqrt{\hat{\rho}_B} \right) \right]^2 
{\rm Tr}({\hat{\rho}}^{\prime} {\hat{\rho}}^{\prime \prime} )>0.
\label{AB}
\end{align}
Equation\ (\ref{AB}) displays the general inequality

\begin{align}
{\cal F}(\hat{\rho}^{\prime}, \hat{\rho}^{\prime \prime}) \geqq
{\rm Tr}({\hat{\rho}}^{\prime} {\hat{\rho}}^{\prime \prime} ),
\label{F>O}
\end{align}
as well as its saturation, which is achieved if and only if the state $\hat{\rho}_B$ is pure:

\begin{align}
{\cal F}(\hat{\rho}^{\prime}, \hat{\rho}^{\prime \prime})=
{\rm Tr}({\hat{\rho}}^{\prime} {\hat{\rho}}^{\prime \prime} ) \;   \iff     \;  
{\rm Tr}\left[ \left( {\hat{\rho}_B} \right)^2  \right]=1.
\label{F=O}
\end{align}
Owing to the formulae\ (\ref{invar}), the purity condition\ (\ref{F=O}) reads

\begin{equation}
{\cal F}(\hat{\rho}^{\prime}, \hat{\rho}^{\prime \prime})=
{\rm Tr}({\hat{\rho}}^{\prime} {\hat{\rho}}^{\prime \prime} ) \;   \iff  \;  \Gamma=\Delta,
\label{G=D}
\end{equation}
and implies the equation

\begin{equation}
\det \left({\mathcal V}_B +\frac{i}{2}J \right)=0 \;   \iff   \;  {\Lambda}=0.
\label{edge}
\end{equation}
The necessary condition\ (\ref{edge}) signifies that at least one of the GSs,
 for instance $\hat{\rho}^{\prime}$, is at the physicality edge: 
$\det \left({\mathcal V}^{\prime}+\frac{i}{2}J \right)=0$. Specifically, 
in the single-mode case $(n=1)$, the corresponding state $\hat{\rho}^{\prime}$ is pure.

Conversely, if one of the above $n$-mode GSs, say $\hat{\rho}^{\prime}$, is pure, 
then so is the GS $\hat{\rho}_B$,  Eq.\ (\ref{rhoB}). Indeed, the required equality 
$\Gamma=\Delta$ is a consequence of the assumed purity conditions 
$\det \left({\mathcal V}^{\prime} \right)=2^{-2n}$ and 
${\mathcal V}^{\prime}=-\frac{1}{4}J \left({\mathcal V}^{\prime} \right)^{-1}J$  \cite{PT2012}.
As shown in Sec. II, this sufficient condition for the saturation property 
${\cal F}(\hat{\rho}^{\prime}, \hat{\rho}^{\prime \prime})=
{\rm Tr}({\hat{\rho}}^{\prime} {\hat{\rho}}^{\prime \prime} )$ is a general one. 
We stress that, for single-mode GSs, it is both necessary and sufficient. 

However, the one-mode case can readily be handled by making direct use 
of the explicit fidelity formula which is available for a long time \cite{HS1998}: 

\begin{align}
{\cal F}({\hat{\rho}}^{\prime}, {\hat{\rho}}^{\prime \prime})
& =\left( \sqrt{\Delta+\Lambda}-\sqrt{\Lambda} \right)^{-1}  \notag \\
& \times \exp{\left[-\frac{1}{2}\left(\delta v \right)^T 
\left({\mathcal V}^{\prime}+{\mathcal V}^{\prime\prime}\right)^{-1} 
\delta v \right]}. 
\label{F1}
\end{align}
Equation\ (\ref{AB}) has the specific form

\begin{align}
{\cal F}(\hat{\rho}^{\prime}, \hat{\rho}^{\prime \prime})
=\left( \sqrt{1+\frac{\Lambda}{\Delta}} +\sqrt{\frac{\Lambda}{\Delta}}\right)
{\rm Tr}({\hat{\rho}}^{\prime} {\hat{\rho}}^{\prime \prime} )>0.
\label{fo1}
\end{align}
Accordingly, the general inequality\ (\ref{F>O}) is manifest and so is the saturation 
condition

\begin{equation}
{\cal F}(\hat{\rho}^{\prime}, \hat{\rho}^{\prime \prime})=
{\rm Tr}({\hat{\rho}}^{\prime} {\hat{\rho}}^{\prime \prime} ) \;   \iff    \;  \Lambda =0,
\label{pure1}
\end{equation}
meaning that at least one of the single-mode GSs $\hat{\rho}^{\prime}$ 
and $\hat{\rho}^{\prime \prime}$ is pure.

\section{Fidelity between thermal states:
The inequality ${\cal F}({\hat{\rho}}^{\prime}, {\hat{\rho}}^{\prime \prime}) \leqq  1$}

Let us introduce the positive function
\begin{align}
& {\cal Q}(x,y):=\sqrt{(x+1)(y+1)} -\sqrt{xy}\, ,    \notag \\
& \left( x \geqq 0, \;\;  y \geqq 0 \right).    
\label{Q}
\end{align}
Remark that the equivalent inequalities 
\begin{equation}
{\left( \sqrt{x}-\sqrt{y} \right)}^2   \geqq 0 \;  \iff   \;   {\cal Q}(x,y) \geqq 1
\label{Q>1}
\end{equation}
become saturate if and only if $x=y$:
\begin{equation}
{\cal Q}(x,y)=1 \;  \iff \;  x=y.
\label{Q=1}
\end{equation}

\subsection{Single-mode thermal states}

A single-mode TS, $\hat{\rho}_{\rm T}(\bar{n})$, is an unshifted  GS whose explicit expression 
is written in Eq.\ (\ref{TS}). Recall that its $2 \times 2$ CM is a multiple of the identity:
\begin{equation}
{\mathcal V}_{\rm T}(\bar{n})= \left( \bar{n}+\frac{1}{2} \right) {\sigma}_0. 
\label{VT1}
\end{equation} 
The fidelity\ (\ref{F1}) of a pair of one-mode TSs, $\hat{\rho}_{\rm T}(\bar{n}^{\prime}) $
and  $\hat{\rho}_{\rm T}(\bar{n}^{\prime \prime})$, is therefore

\begin{equation}
{\cal F}{\left[ \hat{\rho}_{\rm T}(\bar{n}^{\prime}), \, \hat{\rho}_{\rm T}(\bar{n}^{\prime \prime}) \right] }
=\left( \sqrt{\Delta+\Lambda}-\sqrt{\Lambda} \right)^{-1} 
\label{FTS1}
\end{equation}
with the determinants:
\begin{equation}
\Delta =\left( \bar{n}^{\prime}+\bar{n}^{\prime \prime}+1 \right)^2,     \quad
\Lambda=4 \bar{n}^{\prime}(\bar{n}^{\prime}+1)\, \bar{n}^{\prime \prime}(\bar{n}^{\prime \prime}+1).
\label{DL}
\end{equation} 
From Eqs.\ (\ref{DL}) and\ (\ref{Q}) we get the identity
\begin{equation}
\sqrt{\Delta+\Lambda}-\sqrt{\Lambda}={\left[ {\cal Q}{ \left( \bar{n}^{\prime},  
\bar{n}^{\prime \prime} \right) } \right] }^2 
\label{DLQ}
\end{equation}
and the fidelity\ (\ref{FTS1}) reads thereby explicitly: 
\begin{equation}
{\cal F}{\left[ \hat{\rho}_{\rm T}(\bar{n}^{\prime}), \, \hat{\rho}_{\rm T}(\bar{n}^{\prime \prime}) \right] }
=\left[ {\cal Q}{ \left( \bar{n}^{\prime}, \bar{n}^{\prime \prime} \right) } \right]^{-2}. 
\label{FT1}
\end{equation}
Accordingly,  Eq.\ (\ref{Q>1}) displays the inequality
\begin{equation}
{\cal F}{\left[ \hat{\rho}_{\rm T}(\bar{n}^{\prime}), \, \hat{\rho}_{\rm T}(\bar{n}^{\prime \prime}) \right] }
\leqq 1,
\label{FT1<1}
\end{equation}
while Eq.\ (\ref{Q=1}) ascertains its saturation:
\begin{equation}
{\cal F}{\left[ \hat{\rho}_{\rm T}(\bar{n}^{\prime}), \, \hat{\rho}_{\rm T}(\bar{n}^{\prime \prime}) \right] }=1
 \;  \iff   \;  \bar{n}^{\prime}=\bar{n}^{\prime \prime}.
\label{FT1=1}
\end{equation} 

\subsection{Two-mode thermal states}

Let $K_{\pm}(\{\bar{n}\})$ designate the functions\ (\ref{Kpm}) for a pair of two-mode TSs.
They are limit cases of the similar functions for both corresponding pairs of MTSs and STSs. 
In order to write them, it is sufficient to set either ${\theta}^{\prime}={\theta}^{\prime \prime}, \, 
{\phi}^{\prime}={\phi}^{\prime \prime}$ in Eq.\ (\ref{K(-)M}) 
or $r^{\prime}=r^{\prime \prime}, \, {\phi}^{\prime}={\phi}^{\prime \prime}$ 
in Eq.\ (\ref{K(+)S}):
\begin{align}
& K_{+}(\{\bar{n}\})=2\left\{ \left(  \bar{n}_1^{\prime} \, \bar{n}_2^{\prime} \,
 \bar{n}_1^{\prime \prime} \, \bar{n}_2^{\prime \prime} \, \right)^{\frac{1}{2}} \right.  \notag \\
& \left. +\left[ \left(  \bar{n}_1^{\prime}+1 \right) \left( \bar{n}_2^{\prime}+1 \right)
\left( \bar{n}_1^{\prime \prime}+1 \right) \left( \bar{n}_2^{\prime \prime}+1 \right) \right]^{\frac{1}{2}} 
\right\}^2.   
\label{K(+)T}
\end{align}
\begin{align} 
& K_{-}(\{\bar{n}\})=2\left\{ \left[  \bar{n}_1^{\prime} \left( \bar{n}_2^{\prime} +1 \right) 
\bar{n}_1^{\prime \prime} \left( \bar{n}_2^{\prime \prime} +1 \right) \right]^{\frac{1}{2}} \right.  \notag \\
& \left. +\left[ \left(  \bar{n}_1^{\prime}+1 \right) \bar{n}_2^{\prime}
\left( \bar{n}_1^{\prime \prime}+1 \right)  \bar{n}_2^{\prime \prime} \right]^{\frac{1}{2}} 
\right\}^2.   
\label{K(-)T}
\end{align}
The difference between the square roots of the above functions factors as follows:
\begin{align}
& \sqrt{ K_{+}(\{\bar{n}\}) }-\sqrt{ K_{-}(\{\bar{n}\}) }  \notag \\
& =\sqrt{2}\, {\cal Q}{ \left( \bar{n}_1^{\prime}, \bar{n}_1^{\prime \prime} \right) }
\, {\cal Q}{ \left( \bar{n}_2^{\prime}, \bar{n}_2^{\prime \prime} \right) }.
\label{dif}
\end{align}
Substitution of Eq.\ (\ref{dif}) into Eq.\ (\ref{F2(K)}) gives the fidelity of a pair of two-mode TSs:
\begin{align}
& {\cal F}{\left[ \hat{\rho}_{\rm T}{ \left( \bar{n}_1^{\prime}, \bar{n}_2^{\prime} \right) }, \, 
\hat{\rho}_{\rm T}{ \left( \bar{n}_1^{\prime \prime}, \bar{n}_2^{\prime \prime} \right) }\right] }    \notag \\
& =\left[ {\cal Q}{ \left( \bar{n}_1^{\prime}, \bar{n}_1^{\prime \prime} \right) } \,  
{\cal Q}{ \left( \bar{n}_2^{\prime}, \bar{n}_2^{\prime \prime} \right) }  \right]^{-2}. 
\label{FT2}
\end{align}
Equation\ (\ref{Q>1}) confirms therefore the inequality 
\begin{equation}
{\cal F}{\left[ \hat{\rho}_{\rm T}{ \left( \bar{n}_1^{\prime}, \bar{n}_2^{\prime} \right) }, \, 
\hat{\rho}_{\rm T}{ \left( \bar{n}_1^{\prime \prime}, \bar{n}_2^{\prime \prime} \right) }\right] } \leqq 1,   
\label{FT2<1}
\end{equation}
which saturates as stated by Eq.\ (\ref{Q=1}): 
\begin{align}
& {\cal F}{\left[ \hat{\rho}_{\rm T}{ \left( \bar{n}_1^{\prime},  \bar{n}_2^{\prime} \right), } \, 
\hat{\rho}_{\rm T}{ \left( \bar{n}_1^{\prime \prime},  \bar{n}_2^{\prime \prime} \right) } \right] }=1
 \;  \iff   \;  \bar{n}_j^{\prime}=\bar{n}_j^{\prime \prime},    \notag \\
& (j=1,2).
\label{FT2=1}
\end{align}

Taking account of the formula\ (\ref{FT1}) for one-mode TSs, the structure\ (\ref{FT2}) 
of the fidelity between two-mode TSs\ (\ref{TS}) checks the multiplicativity of fidelity 
in this particular case. The reason for which the formulae\ (\ref{FT2})-\ (\ref{FT2=1}) 
can be extended to $n$-mode TSs, regardless of the number of modes, 
is precisely the above-mentioned multiplication rule. 

\section{Fidelity of special two-mode Gaussian states:
The inequality ${\cal F}({\hat{\rho}}^{\prime}, {\hat{\rho}}^{\prime \prime}) \leqq  1$}

In addition to a pair of special two-mode GSs of the same kind, $\hat{\rho}^{\prime}$ 
and $\hat{\rho}^{\prime \prime}$, we envisage the pair of two-mode TSs,
$\hat{\rho}_{\rm T}^{\prime}$ and $\hat{\rho}_{\rm T}^{\prime \prime}$, with the same 
mean thermal photon occupancies: $\{ \bar{n}_1^{\prime},  \bar{n}_2^{\prime} \}$ 
and $\{ \bar{n}_1^{\prime \prime},  \bar{n}_2^{\prime \prime} \}$, respectively.

\subsection{Mode-mixed thermal states}

We define the MTSs $\hat{\rho}_{\rm \, MT}^{\prime}$ and $\hat{\rho}_{\rm \, MT}^{\prime \prime}$
by the sets of their usual parameters, $\{ \bar{n}_1^{\prime}, \, \bar{n}_2^{\prime}, \, {\theta}^{\prime}, 
\, {\phi}^{\prime}\} \,$ and $\{ \bar{n}_1^{\prime \prime}, \, \bar{n}_2^{\prime \prime}, \, {\theta}^{\prime \prime}, \, 
{\phi}^{\prime \prime}\} \,$, respectively.  Inspection of Eqs.\ (\ref{K(+)M})-\ (\ref{K(-)M}) 
and\ (\ref{K(+)T})-\ (\ref{K(-)T}) provides the identities:
\begin{align}
&  K_{+}=K_{+}(\{\bar{n}\}),  \qquad  K_{-}=K_{-}(\{\bar{n}\})    \notag \\
& -\left( \bar{n}_1^{\prime}-\bar{n}_2^{\prime} \right)\, \left( \bar{n}_1^{\prime \prime}
-\bar{n}_2^{\prime \prime} \right) \left\{ 1-\cos{\left( {\theta}^{\prime}-{\theta}^{\prime \prime} \right) }\right.  \notag \\ 
& \left. +\sin{\left( {\theta}^{\prime}  \right) } \sin{\left( {\theta}^{\prime \prime} \right) }
\left[ 1-\cos{\left( {\phi}^{\prime}-{\phi}^{\prime \prime} \right) }\right]  \right\} . 
\label{KMT}
\end{align}
The emerging inequality
\begin{equation}
\sqrt{K_{+}}-\sqrt{K_{-}} \geqq \sqrt{ K_{+}(\{\bar{n}\})}-\sqrt{K_{-}(\{\bar{n}\}) }
\label{KMT1}
\end{equation}
generates via Eq.\ (\ref{F2(K)}) an inequality for fidelities,
\begin{equation}
{\cal F}{\left( \hat{\rho}_{\rm \, MT}^{\prime}, \, \hat{\rho}_{\rm \, MT}^{\prime \prime} \right) } \leqq
{\cal F}{\left[ \hat{\rho}_{\rm T}{ \left( \bar{n}_1^{\prime}, \bar{n}_2^{\prime} \right) }, \, 
\hat{\rho}_{\rm T}{ \left( \bar{n}_1^{\prime \prime}, \bar{n}_2^{\prime \prime} \right) }\right] },
\label{FMT<FT}
\end{equation}
with the saturation condition
\begin{align}
& {\cal F}{\left( \hat{\rho}_{\rm \, MT}^{\prime}, \, \hat{\rho}_{\rm \, MT}^{\prime \prime} \right) }
={\cal F}{\left[ \hat{\rho}_{\rm T}{ \left( \bar{n}_1^{\prime}, \bar{n}_2^{\prime} \right) },        
\hat{\rho}_{\rm T}{ \left( \bar{n}_1^{\prime \prime}, \bar{n}_2^{\prime \prime} \right) }\right] }    \notag \\ 
& \;   \iff   \; {\theta}^{\prime}={\theta}^{\prime \prime}, \, 
{\phi}^{\prime}={\phi}^{\prime \prime}.
\label{FMT=FT}
\end{align}
By use of Eqs.\ (\ref{FMT<FT})-\ (\ref{FMT=FT}) and\ (\ref{FT2<1})-\ (\ref{FT2=1}), 
we get the expected property\ (\ref{F<1}) for MTSs, that is, the inequality
\begin{equation}
{\cal F}{\left( \hat{\rho}_{\rm \, MT}^{\prime}, \, \hat{\rho}_{\rm \, MT}^{\prime \prime} \right) } \leqq 1
\label{FMT<1}
\end{equation}
and its saturation condition as well:
\begin{equation}
{\cal F}{\left( \hat{\rho}_{\rm \, MT}^{\prime}, \, \hat{\rho}_{\rm \, MT}^{\prime \prime} \right) }=1
\;   \iff  \;  \hat{\rho}_{\rm \, MT}^{\prime}=\hat{\rho}_{\rm \, MT}^{\prime \prime}.
\label{FMT=1}
\end{equation}

\subsection{Squeezed thermal states}

Let $\{ \bar{n}_1^{\prime}, \, \bar{n}_2^{\prime}, \, r^{\prime}, \, {\phi}^{\prime} \} \,$ and 
$\{ \bar{n}_1^{\prime \prime}, \, \bar{n}_2^{\prime \prime}, \, r^{\prime \prime}, \, {\phi}^{\prime \prime} \} \,$
be the  parameters of the STSs 
$\hat{\rho}_{\rm \, ST}^{\prime}$ and $\hat{\rho}_{\rm \, ST}^{\prime \prime}$, 
respectively. By looking at Eqs.\ (\ref{K(+)S})-\ (\ref{K(-)S}) and\ (\ref{K(+)T})-\ (\ref{K(-)T}), 
we get the formulae:
\begin{align}
& K_{+}=K_{+}(\{\bar{n}\})     \notag \\ 
& +\left( \bar{n}_1^{\prime}+\bar{n}_2^{\prime}+1 \right) 
\left( \bar{n}_1^{\prime \prime}+\bar{n}_2^{\prime \prime}+1 \right)  
\left\{ \cosh{\left[ 2\left( r^{\prime}-r^{\prime \prime} \right) \right]  -1}\right.  \notag \\ 
& \left. +\sinh{\left( 2r^{\prime}  \right) } \sinh{\left( 2r^{\prime \prime} \right) }
\left[ 1-\cos{\left( {\phi}^{\prime}-{\phi}^{\prime \prime} \right) }\right]  \right\} ,  \notag \\ 
& K_{-}=K_{-}(\{\bar{n}\})
\label{KST}
\end{align}
The obvious inequality
\begin{equation}
\sqrt{K_{+}}-\sqrt{K_{-}} \geqq \sqrt{ K_{+}(\{\bar{n}\})}-\sqrt{K_{-}(\{\bar{n}\}) }
\label{KST1}
\end{equation}
gives rise, via Eq.\ (\ref{F2(K)}), to an inequality for fidelities,
\begin{equation}
{\cal F}{\left( \hat{\rho}_{\rm \, ST}^{\prime}, \, \hat{\rho}_{\rm \, ST}^{\prime \prime} \right) } \leqq
{\cal F}{\left[ \hat{\rho}_{\rm T}{ \left( \bar{n}_1^{\prime}, \bar{n}_2^{\prime} \right) }, \, 
\hat{\rho}_{\rm T}{ \left( \bar{n}_1^{\prime \prime}, \bar{n}_2^{\prime \prime} \right) }\right] },
\label{FST<FT}
\end{equation}
with the saturation case
\begin{align}
& {\cal F}{\left( \hat{\rho}_{\rm \, ST}^{\prime}, \, \hat{\rho}_{\rm \, ST}^{\prime \prime} \right) }
={\cal F}{\left[ \hat{\rho}_{\rm T}{ \left( \bar{n}_1^{\prime}, \bar{n}_2^{\prime} \right) },        
\hat{\rho}_{\rm T}{ \left( \bar{n}_1^{\prime \prime}, \bar{n}_2^{\prime \prime} \right) }\right] }    \notag \\ 
& \;   \iff      \; r^{\prime}=r^{\prime \prime}, \, {\phi}^{\prime}={\phi}^{\prime \prime}.
\label{FST=FT}
\end{align}
By making combined use of Eqs.\ (\ref{FST<FT})-\ (\ref{FST=FT}) together 
with Eqs.\ (\ref{FT2<1})-\ (\ref{FT2=1}), we get the expected property\ (\ref{F<1}) for STSs,
consisting of the inequality 
\begin{equation}
{\cal F}{\left( \hat{\rho}_{\rm \, ST}^{\prime}, \, \hat{\rho}_{\rm \, ST}^{\prime \prime} \right) } \leqq 1
\label{FST<1}
\end{equation}
and its saturation case:
\begin{equation}
{\cal F}{\left( \hat{\rho}_{\rm \, ST}^{\prime}, \, \hat{\rho}_{\rm \, ST}^{\prime \prime} \right) }=1
\;   \iff     \;  \hat{\rho}_{\rm \, ST}^{\prime}=\hat{\rho}_{\rm \, ST}^{\prime \prime}.
\label{FST=1}
\end{equation}

{\bf Acknowledgments}
This work was supported by the Romanian National Authority for Scientific Research, CNCS-UEFISCDI, through Project PN-II-ID-PCE-2011-3-1012 for the University of Bucharest.


\begin{thebibliography}{60}
\bibitem{K} The corresponding principal curvatures at the saddle point $S$ are $k_1  \approxeq -4.19$ 
and $k_2  \approxeq 10.09$. Their product, $K:=k_1 k_2  \approxeq -42.31$, is the Gaussian curvature 
of the surface at its saddle point $S$.
\end{thebibliography}
\end{document}